\documentclass[12pt,preprint]{aastex}

\slugcomment{accepted for publication in ApJ}

\shorttitle{Extended Red Emission in the ``Evil Eye'' Galaxy (NGC4826)}
\shortauthors{Pierini et al.}

\begin{document}

\title{Extended Red Emission in the ``Evil Eye'' Galaxy (NGC4826)}

\author{D. Pierini}
\affil{Ritter Astrophysical Research Center, The University of Toledo,
Toledo, OH 43606}
\email{pierini@ancona.astro.utoledo.edu}

\author{A. Majeed\altaffilmark{1}}
\affil{Dept. of Physics and Astronomy, The University of Toledo,
Toledo, OH 43606}

\author{T. A. Boroson\altaffilmark{2}}
\affil{National Optical Astronomy Observatories,
P.O. Box 26732, Tucson, AZ 85726}
\email{tyb@noao.edu}

\and

\author{A. N. Witt}
\affil{Ritter Astrophysical Research Center, The University of Toledo,
Toledo, OH 43606}
\email{awitt@dusty.astro.utoledo.edu}

\altaffiltext{1}{exchange student from the University of Hertfordshire, UK}
\altaffiltext{2}{Visiting Astronomer, Kitt Peak National
Observatory, National Optical Astronomy Observatory, which is operated by
the Association of Universities for Research in Astronomy, Inc. (AURA)
under cooperative agreement with the National Science Foundation.}

\begin{abstract}
NGC4826 (M64) is a nearby Sab galaxy with an outstanding, absorbing dust lane
(called the ``Evil Eye'') asymmetrically placed across its prominent bulge.
In addition, its central region is associated with several regions
of ongoing star formation activity.
We obtained accurate low-resolution ($\rm 4.3~\AA$/pixel) long-slit
spectroscopy (KPNO 4-Meter) of NGC4826 in the 5300 -- $\rm 9100~\AA$
spectral range, with a slit of 4.4 arcmin length, encompassing the galaxy's
bulge size, positioned across its nucleus.
The wavelength-dependent effects of absorption and scattering by the dust
in the Evil Eye are evident when comparing the observed stellar spectral
energy distributions (SEDs) of pairs of positions symmetrically located
with respect to the nucleus, one on the dust lane side and one on
the symmetrically opposite side of the bulge, under the assumption that
the intrinsic (i.e. unobscured) radiation field is to first order
axi-symmetric.
We analyzed the SED ratios for a given number of pairs of positions through
the multiple-scattering radiative transfer model of Witt \& Gordon.
As a main result, we discovered strong residual Extended Red Emission (ERE)
from a region of the Evil Eye within a projected distance of about 13 arcsec
from the nucleus, adjacent to a broad, bright HII region,
intercepted by the spectrograph slit.
ERE is an established phenomenon well-covered in the literature
and interpreted as originating from photoluminescence by nanometer-sized
clusters, illuminated by UV/optical photons of the local radiation field.
In the innermost part of the Evil Eye, the ERE band extends
from about 5700 to $\rm 9100~\AA$, with an estimated peak intensity
of $\rm \sim 3.7 \times 10^{-6}~ergs~s^{-1}~\AA^{-1}~cm^{-2}~sr^{-1}$
near $\rm 8300~\AA$ and with an ERE-to-scattered light band-integrated
intensity ratio, $I(ERE)/I(sca)$, of about 0.7.
At farther distances, approaching the broad, bright HII region,
the ERE band and peak intensity shift toward longer wavelengths,
while the ERE band-integrated intensity, $I(ERE)$, diminishes
and, eventually, vanishes at the inner edge of this HII region.
The radial variation of $I(ERE)$ and $I(ERE)/I(sca)$ does not match
that of the optical depth of the model derived for the dust lane.
By contrast, the radial variation of $I(ERE)$, $I(ERE)/I(sca)$
and of the ERE spectral domain seems to depend strongly on the strength
and hardness of the illuminating radiation field.
In fact, $I(ERE)$ and $I(ERE)/I(sca)$ diminish and the ERE band shifts
toward longer wavelengths when both the total integrated Lyman continuum
photon rate, $Q(H^0)_{TOT}$, and the characteristic effective temperature,
$T_{eff}$, of the illuminating OB-stars increase.
$Q(H^0)_{TOT}$ and $T_{eff}$ are estimated from the extinction-corrected
$\rm H \alpha~(\lambda = 6563~\AA)$ line intensity and line intensity ratios
$[NII] (\lambda 6583)/H \alpha$
and $[SII] (\lambda \lambda 6716+6731)/H \alpha$, respectively,
and are consistent with model and observed values typical of OB-associations.
Unfortunately, we do not have data shortward of $\rm 5300~\AA$,
so that the census of the UV/optical flux is incomplete.
The complex radial variation of the ERE peak intensity and peak wavelength,
of $I(ERE)$ and $I(ERE)/I(sca)$ with optical depth and strength
of the UV/optical radiation field is reproduced in a consistent way
through the theoretical interpretation of the photophysics
of the ERE carrier by Smith \& Witt, which attributes
a key-role to the experimentally established recognition
that photoionization quenches the luminescence of nanoparticles.
When examined within the context of ERE observations in the diffuse ISM
of our Galaxy and in a variety of other dusty environments such as
reflection nebulae, planetary nebulae and the Orion Nebula, we conclude that
the ERE photon conversion efficiency in NGC4826 is as high as found
elsewhere, but that the size of the actively luminescing nanoparticles
in NGC4826 is about twice as large as those thought to exist
in the diffuse ISM of our Galaxy.
\end{abstract}


\section{Introduction}
The ``Evil Eye'' Galaxy (NGC4826; M64) is a relatively isolated, nearby
Sab galaxy, located at RA(B1950.0) $\rm 12h~54m~16.4s$
and Dec(B1950.0) $\rm +21d~57m~05s$, according to de Vaucouleurs et al.
(1991 - RC3).
Its distance from us is uncertain by a factor of 2:
e.g. Rubin (1994) uses a rounded 5 Mpc distance, as a compromise between
the quoted low and high distances (Braun, Walterbos \& Kennicutt 1992;
Tully 1988; Kraan-Korteweg \& Tammann 1980).
Hereafter we adopt the distance quoted by Rubin.
NGC4826 has an optical size of $\rm 10.0 \times 5.4~arcmin^2$ (RC3)
and position angle of the line of nodes (PA for short)
of $\rm 115^o$, on the basis of the outer contours on the POSS
(Nilson 1973; see also Rubin 1994).
According to the latter author, the red isophotal diameters imply
an inclination of $\rm 60^o$.
We refer the reader to Rubin (1994) for further details about morphology
and geometry of the Evil Eye Galaxy.

NGC4826 owes its surname to the presence of an outstanding, absorbing
dust lane (the Evil Eye), asymmetrically placed across its prominent bulge.
The central region of NGC4826 shows a further peculiarity, since it is
associated with several HII regions (J. Young 2000, private communication;
see also Keel 1983a; Rubin 1994).
At the Evil Eye, NICMOS imaging in the $\rm Pa \alpha$
($\rm \lambda = 1.8751~\mu m$) line, obtained by B\"oker et al. (1999),
reveals the presence of an inner arc-shaped region of ongoing
star formation activity and of a broad lane of widely distributed,
bright HII regions as well, in agreement with the morphology of
the $\rm H \alpha$ ($\rm \lambda = 6563~\AA$) line emission
(J. Young 2000, private communication; see also Keel 1983a; Rubin 1994).
Spectroscopic studies (e.g. Keel 1983b; Alonso-Herrero et al. 2000) classify
this galaxy as a starburst-powered LINER.

Sakamoto et al. (1999) imaged the central region of NGC4826
in the $\rm ^{12} CO (J = 1 - 0)~(\lambda = 2.6~mm)$ line
at the Nobeyama Millimeter Array, with a $\rm 4.6 \times 4.3~arcsec^2$
synthesized beam and an $\rm 8.1~km~s^{-1}$ channel width.
They find that the average hydrogen column density of the molecular clouds
at the Evil Eye is about $\rm 1-3 \times 10^2~M_{\odot}~pc^{-2}$,
2 times larger than on the opposite side of the nucleus,
by adopting the conventional CO-to-$\rm H_2$ conversion factor
$\rm X = 3.0 \times 10^{20}~cm^{-2}~(K~km~s^{-1})^{-1}$, determined
in the Galactic disk (Scoville et al. 1987; Solomon et al. 1987;
Strong et al. 1988).
This average hydrogen column density produces a V-band
($\rm \lambda = 0.55~\mu m$) extinction of $\rm A_V = 7$ -- 21 mag,
if the dust is uniformly distributed and a Galactic dust-to-gas ratio
is assumed, but the net asymmetry in extinction, $\rm \Delta A_V$, due to
the excess gas in the Evil Eye, is only of about 3 -- 10 mag
(Sakamoto et al. 1999).
As noted by the latter authors, significant deviations of the CO-to-$\rm H_2$
conversion factor from the standard value (e.g. Wilson \& Scoville 1990)
do exist within our Galaxy and in different starburst galaxies,
nuclei of galaxies and low-metallicity galaxies (e.g. Maloney \& Black 1988;
Nakai \& Kuno 1995; Israel 1997).
A value of
$\rm X = 1.06~(\pm 0.14) \times 10^{20}~cm^{-2}~(K~km~s^{-1})^{-1}$,
derived by Digel, Hunter \& Mukherjee (1995) in Orion, sounds more reasonable
to us in the case of the dust lane of NGC4826, given the presence of
several HII regions there.
In this case, the CO measurement of Sakamoto et al. (1999) gives
$\rm \Delta A_V = 1$ -- 3.5 mag.

Sakamoto et al. also find that the velocity field is quite ordinary
with a hint of noncircular motions at the radius of four CO clumps,
three of the clumps being associated with the dust lane.
By contrast, two nested counterrotating HI gas disks, at 1 kpc
and in the outer region of the stellar disk, have been discovered
in this galaxy (Braun, Walterbos \& Kennicutt 1992; Rix et al. 1995).
As quoted by Sakamoto et al., the quiescent velocity field suggests
that the Evil Eye does not have much effect on the gas kinematics
in the central kiloparsec (i.e. the central 50 arcsec, assuming
the distance of 4.1 Mpc adopted by these authors).
CO line profiles are single-peaked in the Evil Eye, suggesting
to Sakamoto et al. that most of the molecular gas in the dark lane
has already settled in the disk even if the gas accreted from outside.
This conclusion is at variance with that of Block et al. (1994).
The latter authors imaged NGC4826 in the V-band
and $\rm K^{\prime}$-band ($\rm \lambda = 2.1~\mu m$).
The presence of the Evil Eye produces a striking asymmetry in the V-band
surface brightness profile, the axial symmetry being restored
at $\rm 2.1~\mu m$.
From the analysis of these surface brightness profiles, oriented through
the galaxy's center but sequentially spaced 10 degrees each in position angle,
Block et al. determined a least-squares fit value of
$\Delta V / E(V - K^{\prime}) = 1.04 \pm 0.05$ for the Evil Eye,
by assuming axial symmetry for both the unobscured V-band surface brightness
profile and the unreddened $\rm V - K^{\prime}$ color profile.
The multiple-scattering radiative transfer models of Witt, Thronson \&
Capuano (1992) allowed Block et al. to identify this value with the geometry
of a foreground dusty screen with an average V-band optical depth
$\tau_V \simeq 2.1$.
The latter value is equivalent to $\rm \Delta A_V = 2.3$ mag, consistent
with our previous estimate.
Finally, in the analysis of Block et al., the screen model was also supported
by the value of the far-IR-to-optical luminosity ratio of NGC4826
and by the agreement between computed and measured (molecular$+$atomic)
hydrogen column density.

On the basis of the previous arguments, it is reasonable to assume that
a moderate amount of gas and dust distributed in a foreground screen
is responsible both for the observed amount of attenuation
and for the absence of a remarkable kinematic signature in the velocity field
at the dust lane of NGC4826.
In our opinion, this scenario is supported by the relative weakness
of the average CO surface density in the Evil Eye (Sakamoto et al. 1999),
compared to the high optical opacity there (Block et al. 1994).
In fact, a screen geometry associated with a homogeneous dust distribution
(like in models of Witt, Thronson \& Capuano 1992) gives a lower limit
to the computed mass of the dust responsible for a given (observed) reddening
(e.g. Witt, Thronson \& Capuano 1992).
The clumpy structure of the dust is manifested by the presence of dark
filaments in the NICMOS image of NGC4826 in the $\rm Pa \alpha$ line
(B\"oker et al. 1999).
This confirms previous suggestions on the nature of the gaps present
in the distribution of the $\rm H \alpha$ emission from several HII regions
at the dust lane (J. Young 2000, private communication; Keel 1983a;
Rubin 1994).
This wealth of information makes NGC4826 a suitable test
for radiative transfer models.

The strong effect of a homogeneous vs. clumpy dust distribution
in different star/dust geometries has been highlighted and carefully
investigated recently by the new multiple-scattering radiative transfer
models of Witt \& Gordon (2000 -- hereafter referred to as WG00).
WG00 have computed the fractions of escaping/scattered radiation
in the 1000 -- 30000 $\rm \AA$ spectral range
from three representative types of galactic environments,
filled with either homogeneous or two-phase clumpy dust distributions.
Furthermore, they considered two types of interstellar dust,
similar to those found in the average diffuse interstellar medium (ISM)
of the Milky Way (MW) and in the Bar of the Small Magellanic Cloud (SMC).
For a given optical depth of the model, absolute estimates of
the escaping radiation require the knowledge of the spectral energy
distribution (SED) of the intrinsic (i.e. non attenuated) interstellar
radiation field (ISRF), e.g. through the comparison between data
and calculations of stellar population synthesis models.
In the case of NGC4826, low-resolution spectroscopy of the central region,
with a slit encompassing equally extended regions on the obscured
and unobscured side of the nucleus, provides to first order
an observational determination of the intrinsic stellar SED in the dust lane,
by assuming axial symmetry for the ISRF before dust absorption and scattering
(cf. Block et al. 1994).
As a consequence, it also provides a measurement of the attenuation
of the intrinsic ISRF in the Evil Eye.
A least-squares fit of the computed fractions of transmitted and scattered
radiation at several different wavelengths to the ratio of the observed
and intrinsic SEDs, for a given position across the dust lane, identifies
the corresponding optical depth.

As a further goal, we want to search for a residual dust emission feature,
which manifests itself in the familiar broad 0.7 $\rm \mu m$ band,
known as Extended Red Emission (ERE).
Broad-band ($\rm \Delta \lambda \sim 0.1~\mu m$) ERE,
with a peak intensity wavelength $\lambda_p(ERE)$ between
0.65 and 0.88 $\mu m$, has been reported in many different
dusty astrophysical environments of our Galaxy, e.g. the diffuse ISM
(Gordon, Witt, \& Friedmann 1998), cirrus clouds (Szomoru \& Guhathakurta
1998), reflection nebulae (Witt \& Boroson 1990), planetary nebulae
(Furton \& Witt 1990, 1992), the Orion Nebula (Perrin \& Sivan 1992)
and the high-$b$ dark nebula Lynds 1780 (Chlewicki \& Laureijs 1987;
Mattila 1979).
We refer the reader to e.g. Witt, Gordon \& Furton (1998) for a review
of the observed properties of ERE.
Here, it is sufficient to say that the ERE is commonly interpreted as
a photoluminescence phenomenon originating in dusty environments
illuminated by UV/optical photons.
Identifying ERE in NGC4826 would extend our current knowledge of ERE
to galaxies different from our own, so far limited to the halo of
the starburst galaxy M82 (Perrin, Darbon, \& Sivan 1995) and the 30-Doradus
HII region in the Large Magellanic Cloud (LMC -- Darbon, Perrin \& Sivan
1998), but also provide the theory of the ERE carrier
(Witt, Gordon \& Furton 1998 and references therein) and photophysics
(Smith \& Witt 2001) with a further bench-mark.

We present new optical spectroscopy of NGC4826 in Sect. 2.
Data reduction and analysis are discussed in Sect. 3.
The results of our study are illustrated in Sect. 4.
In Sect. 5 we discuss these results within the context of ERE observations
in different dusty environments of our Galaxy, laboratory studies
of photoluminescence properties of silicon nanocrystals and the models
of Smith \& Witt (2001).
Conclusions are summarized in Sect. 6.

\section{Observations}

The observation of NGC4826 was performed with the Cryogenic Camera,
the low-resolution ($\rm 200 \le R \le 500$) CCD spectrometer in use
at the Mayall 4-Meter Telescope of the KPNO, on the photometric night
of 1 February 1994.
The camera system used a lens collimator, the 780-2 grism
(300 g/mm; 1st order; 4900 -- 9700 $\rm \AA$ spectral range;
$\rm 4.3~\AA$/pixel) and an f/1 classical Schmidt camera with a Ford/Loral
$\rm 800 \times 1200$ chip with $\rm 15~\mu m$ pixels
and a 0.84 arcsec/pixel scale.
This chip is sensitive over the 4000 -- 10000 $\rm \AA$ spectral region,
has a readout noise of $\sim$ 15 -- 18 electrons rms, a full-well
of 200000 electrons and a gain of 1.5 electrons per ADU.
Its quantum efficiency peaks at $\rm \sim 86$\% at $\rm 6500~\AA$.
The overall system efficiency, including the atmosphere, telescope
and instrument, peaks at about 20\%.

The camera system was operated in the long-slit mode, with a slit width
of 2.5 arcsec, resulting in a 4650 -- $\rm 9710~\AA$ spectral range,
with a resolution of $\rm 19~\AA$ (full-width at half-maximum -- FWHM),
as measured from sky lines.
At the distance of 5 Mpc quoted by Rubin (1994) for NGC4826, the slit width
comprises a metric distance equal to 60.6 pc.
The slit was positioned across the galaxy's nucleus, at a position angle
of $\rm 45^o$, $\rm + 20^o$ Eastward from North with respect to the PA
of the minor axis of NGC 4826 (cf. Sect. 1).
With its 4.4 arcmin length, the slit encompassed the galaxy's bulge size
and equal galactocentric distances were covered on the obscured
and unobscured side of the galaxy.
A pictorial representation of the region of NGC4826 intercepted by
the spectrograph slit under investigation is given in Fig. 1.
Here, isophotal contours of the galaxy emission in the $\rm H \alpha$ line
(J. Young 2000, private communication) are superposed onto a grey-scale
two-dimensional representation of the galaxy V-band surface brightness
distribution (R. Evans 2000, private communication).
A cross marks the position of the galaxy center adopted by us,
i.e. RA(B1950.0) $\rm 12h~54m~16.2s$ and Dec(B1950.0) $\rm +21d~57m~10s$.

Five 900 s exposures of NGC4826 were taken at an average air-mass of 1.07,
for a total integration time of 4500 s.
One 80 s exposure of the spectrophotometric standard star Feige 34
was obtained at an average air-mass of 1.02, with the same optical set-up
used for the galaxy, to enable absolute flux calibration.
Finally, one 10 s exposure and one 5 s exposure of a comparison HeNeAr arc
were obtained, with the same optical set-up used for NGC4826, to enable
absolute wavelength calibration of the galaxy and standard spectra,
respectively.

\section{Data reduction and analysis}

\subsection{Data reduction}

Data reduction was performed in the IRAF environment and relied
on the NOAO slit-spectroscopy packages.\footnote{IRAF is the Image Analysis
and Reduction Facility made available to the astronomical community
by the National Optical Astronomy Observatories, which are operated by
AURA, Inc., under contract with the U.S. National Science Foundation.}
All the exposures were bias-subtracted, flat-fielded and flipped,
in order to have wavelengths increasing from left to right
along the CCD rows.
The 5 single exposures of NGC4826 were combined successively
with a median filter.
Correction of the exposures of the target and of the standard
for geometrical distortions in the spatial axis (along CCD columns) 
was performed through two-dimensional mapping of the wavelength
in the raw spectrum of each associated comparison arc,
assumed as the calibration image.
This procedure enabled both the linearization of the raw spectra
in the wavelength direction (i.e. along CCD rows) and their absolute
calibration in wavelength, with an uncertainty of $\rm 0.8~\AA$.
The almost similarly wavelength-calibrated raw spectra of NGC4826
and of Feige 34 were furthermore transformed into
the same absolute wavelength scale, covering the 4650 -- 9710 $\rm \AA$
spectral range.
It was not possible to correct for the systematic increase of the skew
of the line profiles in the lower half of the chip,
which results from some misfunction in the optical set-up of the system.

Both the resultant raw spectra of NGC4826 and of Feige 34
were affected by non-linear tilts in the spatial direction.
These tilts turned highly non-linear in two regions of the chip,
defined approximately by the first and last 150 columns.
Any spectrum requires a correction for its tilt
in the spatial direction before extraction, of course.
This procedure requires the mapping of the maximum intensity
at any given wavelength along the spatial axis.
The case of NGC4826 is particularly complex, since the inner region
of this galaxy showed a strongly asymmetric radial intensity profile
at any wavelength in the 4650 -- 9710 $\rm \AA$ spectral range,
as a consequence of the presence of the dust lane (cf. Block et al. 1994;
Witt et al. 1994).
By contrast, the case of Feige 34 is simple, since its radial intensity
profile is axially symmetric, by definition.
The tilts of the two raw spectra were found to be identical,
when allowing for a slight average displacement between the intensity peaks
of each object at any given wavelength, along the spatial direction.
A careful investigation of the variation of the displacement
between the traces of the two raw spectra at individual wavelengths 
excluded the presence of any trend with wavelength.
However, it also revealed a systematic increase (by a maximum factor of 2)
of the FWHM of the spatial profiles of the target and of the standard
toward the short- and long-wavelength ends of the chip.
In addition to this increase, at random wavelengths within
the short- and long-wavelength ends of the spectrum, the central portion
of the intensity profile of each object shows an irregular shape,
with peak-to-peak variations within $\rm 3 \sigma$ of the rms readout noise.
This irregular shape was already present in the raw spectra before
their linearization in the wavelength direction.
It was more pronounced in the spectral regions where the signal-to-noise ratio
was lower than 10, associated with a particularly poor overall system
efficiency.
The broadening of the FWHM of the spatial profile in these two spectral
regions reduced the maximum value of the observed intensity
and the difference between the observed peak intensity
and the intensities at neighboring positions.
This effect, coupled to photon plus readout noise, produced the random
disappearence of a single, well-defined peak of the observed intensity
profiles of NGC4826 and Feige 34 at the short- and long-wavelength ends
of their spectra.
Reproducing the trace in these two bad regions of the chip required
very high order Legendre polynomials, with the result that spurious features
were introduced in the extracted spectrum of the star.
This result led us to the decision of deleting the regions of baselines
1 -- 150 and 1050 -- 1200 pixels (equivalent to regions
with $\rm \lambda < 5300~\AA$ and with $\rm \lambda > 9100~\AA$, respectively)
in the fitting of the trace of the Feige 34 spectrum.
Consistently, the corresponding spectral regions of the NGC4826 spectrum 
will be deleted in the further data analysis as well.

The Feige 34 spectrum was extracted from a region
of 35 arcsec width along the spatial axis.
By contrast, the NGC4826 spectrum was extracted in a strip
of 4.3 arcmin width across its nucleus, by adopting the trace fit
of the raw Feige 34 spectrum, but allowing for a trace displacement
of about 3 arcsec along the spatial axis.
At each wavelength, the estimate of the trace displacement was derived
from the fitted position of the true maximum intensity of NGC4826,
under the assumption that the intensity distribution
around the central position of the galaxy was represented by a Gaussian.
A three-pixel spatial baseline, equivalent to 2.5 arcsec,
was chosen for this fit.
Larger spatial baselines were found to cause a shift of the new center
toward the unobscured side of the galaxy and overall higher intensities
and a bluer SED of the extracted NGC4826 spectrum at the dust lane side,
with respect to symmetrically opposite  positions at the unobscured side
of the galaxy, in the inner parts of NGC4826.
This behaviour was interpreted as due to the strong asymmetry of
the observed intensity profile of NGC4826 at these scales
(cf. Block et al. 1994; Witt et al. 1994) and motivated
our choice of a three-pixel spatial baseline.

The program spectrum was extracted after background subtraction,
the extraction window being limited by the presence of two regions
of high instrumental noise at the lower and upper extremities of the chip.
Median values of the background at each wavelength were estimated
in two opposite equivalent regions of the raw spectrum,
adjacent to the extraction window, at about 2.1 arcmin radial distance
from the galaxy center, and fitted with a linear relation before subtraction.
At the galaxy center, the background subtracted at e.g.
$\rm \lambda \sim 7500~\AA$ (where no line emission is present) is equal to
$\rm 2.08 \times 10^{-6}~ergs~s^{-1}~\AA^{-1}~cm^{-2}~sr^{-1}$,
with an rms uncertainty of
$\rm 5.24 \times 10^{-8}~ergs~s^{-1}~\AA^{-1}~cm^{-2}~sr^{-1}$.
This value of the background is equal to the mean of the two median values
of the background determined as described before, by construction.
Within the extraction window, the largest variations of the background
subtracted at $\rm \lambda \sim 7500~\AA$ are equal to $\pm$3\%
of the previous value.
They are associated with the extrema on the unobscured
and obscured side of the galaxy, respectively.
The baseline subtracted as background is affected by emission from
the peripheral regions of the galaxy disk, since the slit length is shorter
than the projected angular size of NGC4826 (cf. Sect. 1 and 2).
However, at galactocentric distances larger than 2.1 arcmin,
the NGC4826 disk emission in the optical has a very shallow radial profile
and a negligible intensity, when compared with the emission
in its bulge region (Heraudeau \& Simien 1996).

The intensity profile of NGC4826 at any given wavelength was centered
and registered in the spectrum extraction procedure, so that the galaxy center
is defined as a result.
However, the true galaxy center may actually lie at a slightly different
neighbouring position, as a consequence of the presence of the one-sided
dust lane across the galaxy's nucleus.
Obviously, the centering precision plays a key-role on the robustness
of the following results.
In order to verify the correctness of the NGC4826 center position
at any given wavelength, as given by the spectrum extraction procedure,
we performed the following test.
First, we assumed the center determined in the spectrum extraction procedure
as a reference position and then folded each radial intensity profile
around a fictious center, displaced from this reference position by 0,
$\rm \pm 0.5$ and $\rm \pm 1$ pixel, i.e. 0, $\rm \pm 0.42$
and $\rm \pm 0.84$ arcsec, respectively.
In each case, we extracted radial intensity profiles of NGC4826
at several different wavelengths, and computed the sum of the squares
of the differences between the intensities of the two halves of each profile
at galactocentric distances of 35 -- 80 arcsec, corresponding to regions
beyond the dust lane.
Within this range of absolute radial distances, the galaxy profile
obtained from V-band imaging (Block et al. 1994; Witt et al. 1994)
seems to be to first order axi-symmetric.
We found that the residuals did not depend on wavelength.
Moreover, most of the minimum values of the sums of the squared residuals
were obtained for a displacement of 0 pixel, but a non negligible
fraction of the rest of the minima was associated with a displacement
of 0.5 pixel.
We further analyzed the ratios of the SEDs at symmetrically opposite
positions relative to the NGC4826 center, on the obscured and unobscured
sides of the galaxy, respectively.
We found that, for the 0.5 pixel displacement, these SED ratios showed
a U-downturn at the longest wavelengths, at least in the innermost region
of the galaxy.
This behaviour is at variance with the common understanding of
the wavelength-dependent effects of absorption and scattering by the dust.
Therefore, we concluded that the true galaxy center is possibly shifted,
at maximum, about 0.25 pixels (i.e. 0.21 arcsec) toward the dust lane,
with respect to the reference position.
We discuss the effect of such an uncertainty on our results in Sect. 4.1.

After extraction, the NGC4826 spectrum was corrected for atmospheric
extinction and calibrated in absolute flux by adopting the sensitivity
function and flux conversion determined for Feige 34.
The resultant absolute flux calibration rms error of Feige 34 spectrum
was 1.1\%.
No wiggles in the flux calibration of Feige 34 spectrum were present
within the spectral range of interest.

\subsection{Data analysis}

The NGC4826 spectrum was extracted in the ``split'' format, so that
it was possible to extract further sub-spectra at individual positions,
whose radial distances from the galaxy center were odd integer multiples
of 0.5 pixel (i.e. 0.42 arcsec).
Sub-spectra were extracted within a central region of 50.4 arcsec width,
corresponding to twice the projected angular width of the dust lane
at $\rm PA = 45^o$.

Under the assumption that the intrinsic radiation field of NGC4826
is to first order axially symmetric (Sect. 1), the SED at each position
on the unobscured side of the galaxy, as seen by the observer, once corrected
for Galactic extinction, represents the ISRF at the symmetrically opposite
position on the obscured side, as seen by the dust lane.
Hereafter we neglet the correction for Galactic extinction.
Conversely, the partially obscured SED emerging through the dust lane consists
of partially attenuated, directly transmitted radiation and radiation
from the underlying ISRF scattered into the same beam.
This assumption is equivalent to attributing the observed asymmetry
of the radial intensity profiles of NGC4826 to the presence of the dust lane.
This is supported by the consideration that a much higher degree
of axial symmetry is shown by the $\rm K^{\prime}$-band surface brightness
profile of NGC4826 with respect to the V-band one (Block et al. 1994;
Witt et al. 1994).
With this assumption, the ratios between the observed SEDs
(in the 5300 -- $\rm 9100~\AA$ spectral region) at symmetrically opposite
positions relative to the NGC4826 center, on the obscured and unobscured
side, respectively, represent the fractions of radiation
directly transmitted/scattered by the dust in the dark lane
at different galactocentric distances.
As a consequence of this division, the effects of the Milky Way interstellar
extinction are filtered out.

Blue continuum emission, arising from HII regions in the central region
of the galaxy, associated in particular with the Evil Eye (Fig. 1; see also
Keel 1983a; Rubin 1994; B\"oker et al. 1999), introduces second order
disturbances at the blue-end of the sub-spectra for the corresponding
radial distances, as discussed in Sect. 4.1.
With high confidence, the continuum emission arising from a region
$\rm 20.4 \times 60.6~pc^2$ in size (cf. Sect. 2) associated with any
of these HII regions is contributed by exciting stars and gas
and reprocessed by local dust.
The atomic continuum emission contributed by the gas is the sum of
the two-photon decay emission and of the free-bound continuum radiation.
It is almost constant with wavelength between $\rm \sim 5300~\AA$
and $\rm \sim 8200~\AA$ (the location of the Paschen break)
and between $\rm \sim 8200~\AA$ and $\rm \sim 9100~\AA$, the decrease
at the Paschen break being of a factor of 3 (Perrin \& Sivan 1992;
Sivan \& Perrin 1993; Darbon, Perrin \& Sivan 1998; Darbon et al. 2000).
Hence, the spectral distribution of the atomic continuum emission
does not affect the profile of the observed SED except for the effect
of the Paschen discontinuity, if significant.
Shortward of the Paschen break the atomic continuum emission
also does not exceed 40\% of the optical continuum emission originated
from scattering of starlight by dust grains in a dusty HII region
like the Orion Nebula (Perrin \& Sivan 1992).
This fraction seems to be valid also between $\rm \sim 8200~\AA$
and $\rm \sim 9100~\AA$.
If this estimate applies to HII regions in NGC4826,
the atomic continuum emission is expected not to exceed 17\% of
the total optical continuum emission there, under the assumption that
the stellar contribution is at least equal to that of the scattered light.
A reasonable representation of an HII region environment is the clumpy
SHELL model of WG00, which predicts that, in the red, the scattered light
ranges from 30 to 60\% of the stellar light directly transmitted for
reasonable values of the V-band optical depths, increasing from 1 to 10,
without dependence on the intrinsic opacity function.
As a consequence, in the observed HII region of NGC4826, we may expect that
the atomic continuum emission does not exceed 8 -- 13\% of the total
optical continuum emission.
The non-detection of the Paschen break in the observed spectra
supports the conclusion that the contribution of the atomic continuum
emission to the total optical continuum emission collected by our slit width
is negligible.
Therefore, we do not correct these spectra for the fraction
of the atomic continuum emission before the normalization by
the corresponding unobscured spectra.
In Sect. 4.5 we discuss further evidence for the conclusion that the origin of
the blue continuum emission arising from these HII regions is mostly stellar.

Here we compare the SED ratios with the fractions of
directly transmitted/scattered radiation predicted by the WG00 models
for different astrophysical environments.
For 25 wavelengths, ranging from 1000 to $\rm 30000~\AA$,
WG00 computed the fraction of incoming radiation which is
directly transmitted/scattered by and finally escapes from
a galactic environment filled with either homogeneous or two-phase
clumpy dust distributions.
Each model is identified by its geometry, the intrinsic opacity function
(MW- or SMC-like), the radial extinction optical depth from the center
to the edge of the dust environment in the V-band ($\rm \tau_V$),
assuming a constant density homogeneous distribution,
and the type of structure (homogeneous/clumpy) in the dust region.
Since the detailed analysis of the reddening in the Evil Eye
(Block et al. 1994) and the comparison of the CO map (Sakamoto et al. 1999)
with the $\rm Pa \alpha$ image (B\"oker et al. 1999 ) of NGC4826
suggest that the {\it global} distribution of the dust with respect to
the stars resembles a screen geometry while the {\it local} distribution
of the dust is clumpy, we adopted the WG00 clumpy SHELL model.
Both the MW- and SMC-types of interstellar dust were considered.

We determined a new grid of $\rm \tau_V$-dependent model fractions
at each wavelength by fitting the predicted fractions of incoming radiation
which is directly transmitted/scattered, quoted by WG00 at discrete values
of $\rm \tau_V$, with a $\rm 4^{th}$ order polynomial in $\tau_V$.
After a visual comparison of the SED ratios previously obtained
with the WG00 original fractions, we limited this fitting
to the WG00 original fractions in the 0.25 -- 5 $\tau_V$ range
and computed new model fractions in the 0.01 -- 5 $\tau_V$ range,
with a step of 0.01 in $\tau_V$.
The fitting was limited to the WG00 original fractions
at the four wavelengths $\rm \lambda \lambda$ 4754, 5985, 7535
and $\rm 9487~\AA$, encompassing the spectral range of the NGC4826 spectrum.
The comparison between these new model ratios and the WG00 original ones,
at the same values of $\tau_V$, showed an agreement better than
a few per cent, the largest disagreement being at $\rm 4754~\AA$
for values of $\tau_V$ either $\le 0.25$ or $\ge 4.5$.
For each value of $\tau_V$, in the range 0.01 -- 5, the new model fractions
of incoming radiation, which is directly transmitted/scattered
at the previous four wavelengths, were fitted with a $\rm 2^{nd}$ order
polynomial in $\rm \lambda$, in order to obtain a continuous characterization
of the wavelength-dependent effects of absorption and scattering
within the whole 5300 -- $\rm 9100~\AA$ spectral range.
A $\rm 2^{nd}$ order polynomial was found to reliably fit the original
WG00 model fractions at $\rm \lambda \lambda$ 4754, 5985, 7535
and $\rm 9487~\AA$.

\section{Results}

\subsection{Extinction and scattering within the dark lane of NGC4826}

Fig. 2a--n shows the SED ratios determined in Sect. 3.2 for absolute
projected radial distances from the NGC4826 nucleus in the 2.1 -- 23.9 arcsec
range, with a step of 1.7 arcsec.
We recall that the 0.84 arcsec/pixel angular scale of the CCD used for
the present spectroscopy of NGC4826 corresponds to a 20.4 pc/pixel
metric scale at the distance of this galaxy (cf. Sect. 1).
The SED ratio for the innermost galactocentric distances of 0.4
and 1.3 arcsec have a strongly pronounced small-scale saw-tooth
pattern, which is a high-spatial-frequency residual of the corrections
discussed in Sect. 3.1.
As a consequence, we did not fit the SED ratio for the galactocentric
distance $r = 0.4$ arcsec and this SED ratio is not reproduced in Fig. 2.
A less pronounced small-scale saw-tooth pattern is still present
in several of the ratios displayed in Fig. 2, especially in those
associated with the shortest radial distances.
This behaviour is persistent in the SED ratios determined
in the test procedure described in Sect. 3.1 and, therefore, it is not due
to the centering uncertainty.
Furthermore, Fig. 2 shows evidence for bad sky-lines removal
in well-confined spectral regions, such as those centered at about
$\rm \lambda \lambda$ 6300, 7400 and $\rm 7500~\AA$.

Despite these ``defects'', the SED ratios show a strong wavelength
dependence, for any galactocentric distance.
In fact, in each panel of Fig. 2, the SED ratio increases from the blue end
to the red end of the spectral range.
This general behaviour is expected on the basis of the wavelength-dependent
effects of absorption and scattering by dust.
The overall inspection of Fig. 2 shows further systematic changes
of the SED ratio with $r$, along the whole spectral range.
An initial monotonic decrease of the SED ratio, at any given wavelength,
with $r$ increasing from 2.1 (Fig. 2a) to 8.0 arcsec (cf. Fig. 2b--d),
is followed by almost identical SED ratios, along the whole spectral range,
within the 8.0 -- 13.9 arcsec radial range (cf. Fig. 2e--h), while
the SED ratios increase again, along the whole spectral range,
at galactocentric distances greater than 13.9 arcsec (Fig. 2i--n).
This distance-dependent behaviour is easily interpreted as due to
radial changes of the optical depth across the dust lane, as confirmed
by the values of $\tau_V$ associated with our best-fit WG00 model curves
(cf. Sect. 3.2 and Sect. 4.2).

As a further result, we find that the SED ratios for $r \le$ 19.7 arcsec
(cf. Fig. 2a--l) can not be fitted by any WG00 model curve along
their whole spectral range.
In particular, the observed SED ratios exceed the model predictions
in the 5700 -- $\rm 9100~\AA$ spectral range for $r =$ 1.3 arcsec,
this spectral range moving toward longer wavelengths and the excess
diminishing for 1.3 $< r <$ 13.0 arcsec (cf. Fig. 2a--h).
The discrepancy between model curves and data is even more dramatic
if we use MW/SMC extinction curves.
In fact, no SED ratio can be fitted by any of the latter curves.
We show a representative case of this result in Fig. 3,
where we reproduce and zoom the SED ratio for $r =$ 22.3 arcsec together
with its best-fit WG00 model curve with MW-type dust from Fig. 2m
and the MW extinction curve for the same value of $\tau_V$ as the WG00 model,
as tabulated in WG00.
We choose this SED ratio as an example because at the distance of
22.3 arcsec no excess emission (either at the blue- or at the red-end
of the spectral range) is present (cf. Fig. 2a--n).
Fig. 3 shows that while both the WG00 model curve with MW-type dust
and the MW extinction curve fit the data comparably well
at $\rm 5300 \le \lambda \le 6200~\AA$, the MW extinction curve predicts
particularly higher fractions of escaping radiation than those shown by
the data at longer wavelengths.
This is true in particular at $\rm \lambda \ge 8400~\AA$, where, conversely,
the WG00 model curve with MW-type dust fits the data as well as
at the blue-end of the spectral region. 
Fitting the SED ratio at $\rm \lambda \ge 8400~\AA$ with a MW extinction
curve would require a slightly higher optical depth but this new fit would
systematically underestimate the SED ratio
at $\rm 5300 \le \lambda \le 6200~\AA$.
From this comparison, we conclude that the effect of scattering by dust
on the spectral energy distribution of the radiation escaping through
a dusty environment is not negligible and can be treated properly
only via radiative transfer calculations.
As a corollary result, we also conclude that the behaviour in Fig. 2a--h
is not a spurious effect of the WG00 radiative transfer model.
We need to examine whether this behaviour is due either to
an uncertain centering of the NGC4826 spectrum or to the violation
of our assumption of an axi-symmetric intrinsic ISRF.

In order to answer the first question, we determined a new set of SED ratios
under the assumption that the true galaxy center is displaced
by 0.5 pixel toward the dust lane, i.e. twice the estimated centering
uncertainty (cf. Sect. 3.1).
Accordingly, we derived a further set of SED ratios by considering pairs
of symmetrically opposite positions, whose radial distances from the newly
determined galaxy center are even integer multiples of 0.5 pixel
(i.e. of 0.42 arcsec).
We found that (1) at any given galactocentric distance, these SED ratios
were lower than those in Fig. 2, along the whole spectral range,
but that (2) the radial behaviour of these SED ratios was the same
as that displayed in Fig. 2 and that (3) the shapes of these SED ratios
for $r \le$ 19.7 arcsec, although slightly different from those displayed
in Fig. 2, showed a similar excess which could not be fitted either by
any of the WG00 model curves or by any of the MW/SMC extinction curves.
Therefore, we conclude that the behaviour of the SED ratios shown in Fig. 2
for $r \le$ 19.7 arcsec is not due to the centering uncertainty.

The central region of NGC4826 hosts several regions of ongoing star
formation activity, with different extensions and intensities (cf. Fig. 1).
Fig. 2h--l shows that the SED ratios for 13.0 $\le r \le$ 19.7 arcsec
have an almost flat or slightly rising shape in the spectral region
at $\rm \lambda < \sim 7500~\AA$.
This behaviour is ``anomalous'', since it is at variance with
the well-known wavelength-dependent effects of dust absorption and scattering,
under the assumption that the intrinsic ISRF of NGC4826 is axi-symmetric.
We interpret this excess emission at $\rm \lambda < \sim 7500~\AA$
for 13.0 $\le r \le$ 19.7 arcsec as due mainly to stars in a broad, bright
HII region, associated with the Evil Eye, intercepted by the spectrograph
slit, as discussed in Sect. 4.5.
This interpretation is supported by the strength of the residual
line emission for 13.0 $\le r \le$ 19.7 arcsec (Fig. 2h--l)
and is in agreement with the phenomenology of the $\rm Pa \alpha$ line
emission of NGC4826 (B\"oker et al. 1999).
It implies a violation of the axial symmetry assumption for
the intrinsic ISRF of NGC4826 (Sect. 1) and explains the failure
of our fitting procedure along the 5300 -- $\rm 9100~\AA$ spectral range,
for galactocentric distances in the 13.0 -- 19.7 arcsec range.
Minor violations of our assumption are manifested by residuals
in line subtraction and by excess emission
at $\rm \lambda < \sim 5350~\AA$ in the SED ratios for $r \le$ 12.2 arcsec
(cf. Fig. 2a--g).
The SED ratio for $r = 12.2$ arcsec still shows relatively strong
line emission features with respect to inner SED ratios,
so that $r = 12.2$ arcsec may be identified as the inner edge
of the previous broad, bright HII region.
Even deleting the spectral region defined by $\rm \lambda < \sim 7500~\AA$,
the SED ratios for $r \le 12.2$ arcsec can still not be fitted either by any
of the WG00 model curves or by any of the MW/SMC extinction curves.
Therefore, we conclude that this impossibility is significant.
The only physically reasonable interpretation of the behaviour
of the SED ratios for $r \le 12.2$ arcsec is the presence
of a broad featureless band emission arising from the dust lane,
which causes the significantly higher values of these ratios,
with respect to those predicted by the WG00 model curves and by
the MW/SMC extinction curves, longward of $\rm \sim 5700~\AA$.
Given the existence of analogous examples in the literature (Sect. 1),
we interpret this excess red emission, arising from the region
of the dust lane of NGC4826 for $r < 13.0$ arcsec, as due to ERE.

With this tentative interpretation, we fit our model curves
to each SED ratio in the spectral region where neither ERE
or excess blue emission is reasonably found to be significantly present.
In Sect. 4.2, we will compare the properties of the hypothetical ERE
from the dust lane of NGC4826 with the phenomenology of the ERE
reported in the literature, in order to test our interpretation.
As a consequence of the previous discussion, we fit our model curves
to the SED ratios in the 5350 -- $\rm 6000~\AA$ spectral region,
for $r \le 12.2$ arcsec, and in the 7800 -- $\rm 9100~\AA$ spectral region,
for $13.0 \le r \le 21.4$ arcsec, while we adopt an average
of the fits obtained in these two spectral regions
for 22.3 $\le r \le$ 25.6 arcsec.
Slight changes of the limits of these spectral regions introduce variations
of the order of a few per cent on the values of $\lambda_p(ERE)$,
$I_p(ERE)$, $I(ERE)$ and $I(ERE)/I(sca)$, quoted in Sect. 4.2,
due to different best-fit model curves, which are insignificant
for the purposes of the present analysis.

Each least-squares best-fit WG00 model curve identifies a value of $\tau_V$,
where $\tau_V$ is the extinction optical depth of the dust lane
along the sight line in the V-band, assuming a constant density
homogeneous distribution of dust (cf. Sect. 3.2).
From the analysis of the $\chi^2$ values of the least-squares fit model
curves, obtained for each SED ratio, we attribute an uncertainty
of a few per cent to the best-fit $\tau_V$. 
Since $\tau_V$ determines the model fraction of light scattered/directly
transmitted by dust at any given wavelength, it is straightforward
to derive the SEDs of the scattered/directly transmitted light. 
ERE flux densities are derived by subtraction of these contributions
from the flux densities actually observed in the dust lane.
SMC-type dust models require values of $\tau_V$ up to 4\% larger
than the MW-type dust ones, as the HII region is approached,
this discrepancy diminishing when the outer edge of the dust lane is reached.
Given this systematic but slight discrepancy, we conclude that the ERE flux
densities are not affected by which type of dust is adopted.

We note that, taking best-fit $\chi^2$ values at face-value, we find hints
that SMC-type dust describes the attenuation in the core of the prominent
HII region observed in the dust lane better than MW-type dust.
Although this result is consistent with the finding that starburst galaxies
have LMC-like extinction laws (Gordon, Calzetti \& Witt 1997), we do not
claim any statistically robust evidence of this effect.
On the other hand, the Gordon et al.'s conclusion was based
on an analysis in the UV spectral range, where the differences between
the MW and the SMC dust become much more pronounced.
Since there are no remarkable differences between the best fit $\chi^2$
values obtained either with SMC- or with MW-type dust everywhere else
in the NGC4826 bulge region within our spectral range (cf. Fig. 2),
hereafter we show and discuss results obtained under the neutral assumption
that the dark lane of NGC4826 hosts dust of MW-type.

We plot the (assumed) intrinsic SED of the ISRF (as seen by the observer)
and the (calculated) SED of the ISRF escaping through the dust lane,
together with the observed flux density distribution, for different positions
across the dark lane of NGC4826 in the upper portion of Fig. 4a--n.
For the same vaues of $r$, we plot the model spectrum of the light
scattered by dust and the residual ERE intensity profile in the lower portion
of Fig. 4a--n.

Where detected, the ERE intensity at $\rm \lambda \sim 7500~\AA$
ranges from about twice to 15\% of the background subtracted
at this wavelength in the data reduction (cf. Sect. 3.1).
Intensities within this range are significant with respect to
both the rms uncertainty of the background at $\rm \lambda \sim 7500~\AA$
and the maximum variation of this background within the extraction window.
This result extends to the whole spectral range under study.

\subsection{ERE properties in the Evil Eye}

From the visual inspection of Fig. 4, it emerges that (1) the ERE band
and peak intensity shift toward longer wavelengths with increasing
projected galactocentric distance and that (2) the ERE peak intensity
and the ERE band-integrated intensity decrease at the same time.
Band-integrated means that the wavelength domain of integration
corresponds to the wavelength range where ERE is detected
(limited longward at $\rm 9100~\AA$), for each value of $r$.
This behaviour is confirmed more quantitatively by fitting the smoothed
ERE intensity profiles with Legendre functions and by numerical integration
of the same profiles within the wavelength range of the ERE band
(shorter than the 5300 -- $\rm 9100~\AA$ spectral range).
Smoothing was obtained by averaging the ERE intensity in wavelength bins
of $\rm 122.4~\AA$ width.
This width is arbitrary; however, it is a fairly good compromise
in order to reproduce the ERE intensity profile via a discrete
but not coarse sampling, without being biased by residuals
of an uncorrect line-subtraction.
We used the lowest order Legendre function reproducing at best the ERE
intensity profile within the wavelength range of ERE detection,
for any given $r$.
This best-fit may not represent the unknown and unpredictable shape
of the ERE intensity profile of the Evil Eye longward of $\rm 9100~\AA$.
Therefore, it does not provide robust estimates of the total ERE
band-integrated intensity and of the fraction of it represented by
the measured ERE band-integrated intensity.
In particular, for $r = 1.3$ arcsec, ERE is detected
at $\rm \lambda > \sim 5700~\AA$, with a peak intensity
$I_p(ERE) \rm \sim 3.7 \times 10^{-6}~ergs~s^{-1}~\AA^{-1}~cm^{-2}~sr^{-1}$
near $\lambda_p(ERE) \rm \sim 8300~\AA$.
By contrast, for $r = 5.5$ arcsec (Fig. 4c), ERE is detected
at $\rm \lambda > \sim 6200~\AA$, with
$I_p(ERE) \rm \sim 2.3 \times 10^{-6}~ergs~s^{-1}~\AA^{-1}~cm^{-2}~sr^{-1}$,
and $\lambda_p(ERE) \rm \sim 8800~\AA$, while, for $r = 8.8$ arcsec (Fig. 4e),
ERE is detected at $\rm \lambda > \sim 6300~\AA$,
with $\lambda_p(ERE) > 9100~\AA$.
Finally, for $r \ge 13.0$ arcsec, ERE is not detected,
at least within the 5300 -- $\rm 9100~\AA$ spectral range (cf. Fig. 4h--n).
The same smoothing and band-integration procedure was applied
to each model spectrum of the scattered light, so that we obtained
the ERE-to-scattered light band-integrated intensity ratio.

We plot the radial behaviour of the ERE band-integrated (within
the 5300 -- $\rm 9100~\AA$ spectral range) intensity, $I(ERE)$, and of
the ERE-to-scattered light band-integrated ratio, $I(ERE)/I(sca)$,
in Fig. 5a,b, respectively.
Globally, we find that $I(ERE)$ decreases from
$\rm \sim 9.4 \times 10^{-3}~ergs~s^{-1}~cm^{-2}~sr^{-1}$ to 0,
with $r$ increasing from 1.3 to 13.0 arcsec (Fig. 5a).
Consistently, $I(ERE)/I(sca)$ declines from 0.67 to 0 as well,
though not in a monotonic way as for $I(ERE)$ (Fig. 5b).
In fact, while $I(ERE)$ falls steeply for $r \le 4.6$ arcsec,
its decrease becomes more gradual at greater galactocentric distances.
By contrast, after an initial fall from 0.67 to 0.25
for $r \le 4.6$ arcsec, $I(ERE)/I(sca)$ rises again to $\sim 0.4$
at $r = 8.0$ arcsec and then falls toward 0 for radial distances
greater than 11.3 arcsec.
We estimate statistical uncertainties of the order of 7\% for $I(sca)$,
of 9\% for $I(ERE)$ and of 11\% for $I(ERE)/I(sca)$.
Therefore, the radial variation of $I(ERE)$ and $I(ERE)/I(sca)$
is significant.

The shift of the ERE band and peak wavelength toward longer wavelengths
(Fig. 4a--g), as the broad HII region of NGC4826 is approached,
is analogous to those manifested in the transition from low-UV photon density
environments (e.g. Galactic diffuse ISM) to high-UV photon density
environments (e.g. Orion Nebula) (cf. Gordon, Witt \& Friedmann 1998;
Smith \& Witt 2001).
In fact, $\lambda_p(ERE) \rm \sim 6100~\AA$ in the ISM,
while $\lambda_p(ERE) \rm \sim 8200~\AA$ in the Orion Nebula bar
adjacent to the Trapezium stars, where the radiation density
is several orders of magnitude higher.
For the latter object, $I(ERE)$ ranges between $\sim 7 \times 10^{-3}$
and $\rm \sim 5 \times 10^{-2}~ergs~s^{-1}~cm^{-2}~sr^{-1}$,
and $I(ERE)/I(sca)$ ranges from 0.2 to 0.6 (Gordon, Witt \& Friedmann 1998).
These ranges are in agreement with those spanned by $I(ERE)$
and $I(ERE)/I(sca)$ across the Evil Eye (Fig. 5a,b).
This comparison supports our hypothesis about the nature of the excess
red emission discovered in the comparison of the SED for pairs
of symmetrically opposite positions, across the NGC4826 nucleus,
on the obscured and unobscured side of the galaxy (Sect. 4.1).

This conclusion is not biased by systematic errors.
We estimate that the presence of extra blue emission for $r \le 4.6$ arcsec
(Sect. 4.1) may cause a systematic overestimate of the corresponding best-fit
model values of $\tau_V$ of the order of a few per cent.
As a consequence, the ensuing underestimate of $I_p(ERE)$, $I(ERE)$
and $I(ERE)/I(sca)$ is negligible.
Potentially more severe is the effect of the uncertainty in centering.
In the extreme and unrealistic case that the true galaxy center
is shifted by 0.42 arcsec (cf. Sect. 3.1), we find an average increase
of $\tau_V$ up to 0.4, so that $I(ERE)$ is reduced to one third
of its previous value, while $I(ERE)/I(sca)$ is less affected.
This value corresponds to an extremely conservative upper limit
to the systematic error of $I(ERE)$.

\subsection{ERE and dust column density in the Evil Eye}

As mentioned in Sect. 1, the ERE phenomenon requires the presence of
both dust and UV/optical photons.
Here, we inspect the radial variation of the ERE integrated properties
and of the dust column density, as measured by the optical depth,
in the Evil Eye.

We plot the radial dependence of the attenuation optical depth
in the V-band, $\tau_V(att)$, in Fig. 5c, together with that
of $I(ERE)$ and $I(ERE)/I(sca)$.
The attenuation optical depth $\tau_{\lambda} (att)$ is defined by WG00 as:
\[\rm \tau_{\lambda} (att) = -ln~f_{\lambda} (esc),\]
where $f_{\lambda} (esc)$ is the fraction of the radiation flux density
at wavelength $\lambda$ (in our case at the V-band effective wavelength)
escaping through the dust lane.
On average, $\tau_V(att)$ is about 30 -- 50\% of $\tau_V$, the extinction
optical depth of the dust lane along the sight line in the V-band,
determined by assuming a constant density homogeneous distribution of dust.
The lower value of $\tau_V(att)$ compared to that of $\tau_V$ is due,
in part, to the reduced opacity of a clumpy medium, and, in part, to
the partial return of radiation scattered into the beam by dust.
However, $\tau_V$ remains a useful parameter in that it measures
the actual average dust column density.
Fig. 5c shows that the maximum $\tau_V(att)$ is about 1.6,
equivalent to a maximum $\tau_V \sim 4.3$.
The mean $\tau_V(att)$ is $\sim 1.1$, equivalent to a mean $\tau_V \sim 2.5$.
If the latter value is held as representative of the whole dust lane
of NGC4826, then it is consistent with the average V-band optical depth
of $\sim 2.1$ determined by Block et al. (1994).
It also favours a low value of the hydrogen column density
at the Evil Eye (cf. Sect. 1).

A range of values of $\tau_V(att)$ are found to exist across the dust lane.
For $r \le 4.6$ arcsec, $\tau_V(att)$ rises steeply from 0.16 to 1.06
with increasing $r$.
For $4.6 < r \le 8.0$ arcsec, $\tau_V(att)$ increases up to $\sim 1.6$
and holds this maximum value for $8.0 \le r \le 13.9$ arcsec, while,
for $13.9 < r \le 18.9$ arcsec, $\tau_V(att)$ drops to $\sim 0.8$. 
For greater distances, approaching the exit from the dark lane,
$\tau_V(att)$ decreases smoothly toward 0.
We note that the less certain interpolation/extrapolation
of the WG00 calculations at/to values of $\tau_V < 0.25$ (cf. Sect. 3.2)
does not affect this behaviour.

By contrast, $I(ERE)$ and $I(ERE)/I(sca)$ decrease toward 0,
for $r$ increasing from 1.3 to 13.0 arcsec, and, in particular,
their radial variation is at variance with that of $\tau_V(att)$
for $r \le 4.6$ arcsec (Fig. 5a,b).
Therefore, we conclude that the radial dependence of $I(ERE)$
and $I(ERE)/I(sca)$ manifested across the Evil Eye can not be simply
attributed to a radial change of the dust column density.
Instead, we infer that the ERE mechanism is increasingly quenched
when the UV-rich radiation environment of the HII region is approached.

\subsection{The integrated Lyman continuum photon flux across the Evil Eye}

The conclusions reached in Sect. 4.2 and 4.3 and the observation that
both $I(ERE)$ and $I(ERE)/I(sca)$ diminish in direction of a broad,
bright HII region associated with the dark lane, intercepted by
the spectrograph slit, suggest that the UV/optical component
of the local ISRF not only powers the ERE but also regulates
the photophysics of its carrier (Smith \& Witt 2001).

Unfortunately, our data do not characterize the stellar optical flux
at $\rm \lambda < 5300~\AA$.
However, we can estimate the strength and the hardness of the UV emission
as follows.
Fig. 6 reproduces the radial intensity profile of NGC4826
at the rest-frame effective wavelength of the R-band
($\rm \lambda = 7000~\AA$ -- short-dashed line) and at the rest-frame
$\rm H \alpha$ line central wavelength
($\rm \lambda = 6563~\AA$ -- solid line), as measured from our
long-slit spectrum.
The latter intensity is not corrected for the contribution of
the neighbouring [NII] ($\rm \lambda \lambda =$ 6548, $\rm 6583~\AA$) lines.
First, we note that Fig. 6 confirms the global axial symmetry
of these two profiles at radial distances greater than $\sim$ 35 arcsec,
beyond the dust lane (cf. Sect. 3.1).
Furthermore, and more interestingly, these two radial intensity profiles
are significantly different in three well-defined regions of the galaxy,
identified as regions of massive star formation activity by the relatively
large $\rm \lambda 6563$-to-$\rm \lambda 7000$ intensity
ratio, this ratio being equivalent to first order to the $\rm H \alpha$
line equivalent width (e.g. Rubin 1994).
These three regions of ongoing star formation correspond (1) to the galaxy
nucleus, (2) to a relatively extended region within the 8 -- 13.9 arcsec
galactocentric distance range (visible in the unobscured region,
on the left side of each profile in Fig. 6), and (3) to a broad, bright
HII region associated with the dust lane, with peak emission
at $r \sim$ 13.9 arcsec.

A better description of the strength of the ionizing UV photon field across
the dust lane is given by the integrated $\rm H \alpha$ line intensity,
as derived through fitting via Voigt profiles and deblending
the $\rm H \alpha$ line and the adjacent [NII] doublet.
In Fig. 7, we plot the radial variation of the integrated $\rm H \alpha$
line intensity, $I(H \alpha)$, across the dust lane of NGC4826,
folded around its nucleus.
Here, the solid and long-dashed lines refer to the unobscured
and obscured regions of the galaxy, respectively.
The typical uncertainty of the integrated $\rm H \alpha$ line intensity
is about 10\% or lower.
This figure increases up to about 50\% for the sub-spectra
relative to radial distances of $\rm \pm 1.3$ arcsec, where the intensities
of the two adjacent [NII] lines are at least comparable with $I(H \alpha)$,
and where the underlying continuum is badly estimated.
Despite this large error, our mean central (within a strip of
$\rm 2.5 \times 4.7~arcsec^2$) value of $I(H \alpha)$ is about half of that
quoted by Keel (1983b)
($\rm 5.7 (\pm 0.24) \times 10^{-4}~ergs~s^{-1}~cm^{-2}~sr^{-1}$),
obtained within a larger area, equivalent to a circular aperture
of 4.7 arcsec in diameter.
Given the aspect of the nuclear region of NGC4826 in the $\rm H \alpha$ line
emission (Fig. 1), these two measurements are consistent.
Our higher spatial resolution suggests the existence of a minimum
$\rm H \alpha$ line intensity at the actual center of the galaxy (Fig. 7).
If this is real, the inconsistency with the presence of a maximum rest-frame
$\rm \lambda 6563$ intensity in Fig. 6 is due to contamination
of the latter intensity by strong [NII] line emission.
This explanation is supported by the recognition that NGC4826 is probably
a starburst-powered LINER (cf. Sect. 1), so that its nucleus
is characterized by a high [NII]-to-$\rm H \alpha$ line intensity ratio
(as also found by us -- cf. Fig. 9b).
An uncertainty by a factor of 2 affects values of $I(H \alpha)$ lower than
$\rm \sim 3.4 \times 10^{-5}~erg~s^{-1}~cm^{-2}~sr^{-1}$, corresponding to
very small $\rm H \alpha$ line equivalent widths.
Such low values of $I(H \alpha)$ are found at galactocentric distances
larger than 15 about arcsec and 22 arcsec on the unobscured and obscured side
of the galaxy, respectively.

Globally, the phenomenology of the $\rm H \alpha$ line emission delineated
in Fig. 6 and in Fig. 7, are consistent.
The latter figure, however, reveals the existence of two equally displaced
but un-even $\rm H \alpha$ line intensity peaks, near the galaxy nucleus
(at $r \rm \sim 3$ arcsec), which might be interpreted as features
of a ring-shaped region of massive star formation activity.
The almost 50\% lower $\rm H \alpha$ line intensity peak
occurs on the unobscured side of the galaxy.
The suggested ring-shaped region of $\rm H \alpha$ line emission seems
to be consistent with the arc-shaped morphology of the $\rm Pa \alpha$
emission revealed by NICMOS imaging (B\"oker et al. 1999), when taking
into account the higher spatial resolution (0.2 arcsec/pixel) of NICMOS.
By contrast, this structure is unresolved in the $\rm H \alpha$ image
of Young (2000, private communication) displayed in Fig. 1, where
the peak emission is saturated, and in the $\rm H \alpha$ image
(0.86 arcsec/pixel resolution) of Rubin (1994).
The $\rm H \alpha$ line emission from the hypothetical inner ring,
might be partly associated with the presence either of an AGN
(Ho, Filippenko, \& Sargent 1997) or, more favourably, of a starburst-powered
LINER (Keel 1983b; Alonso-Herrero et al. 2000).
At larger galactocentric distances ($\sim 6.3 < r < 12.3$ arcsec),
on the unobscured side, an extended region of multiple $\rm H \alpha$ line
intensity peaks, comparable in strength with the previous one, is found.
Here, $I(H \alpha)$ is on average about 2.5 times higher than
at the symmetrically opposite side of the galaxy.
At these galactocentric distances, on the obscured side,
$\tau_V(att)$ increases from 1.3 to 1.6 (cf. Fig. 5b),
so that at least about 20\% of the emerging radiation escapes.
In this region of the Evil Eye, we assume that dust almost completely
obscures $\rm H \alpha$ line emission as intense as that observed
on the opposite side of the galaxy, but we also allow for an excess
$\rm H \alpha$ photon flux, as can be inferred from ground-based
$\rm H \alpha$ and HST-NICMOS imaging.
Farthermore, across the region of the Evil Eye probed by our spectroscopy,
a very bright and relatively extended HII region is encountered,
which has no counterpart at the symmetrically opposite position,
on the unobscured side of the NGC4826 bulge region.
The presence of this broad, bright HII region, associated with the dust lane,
violates the assumed axial symmetry of the intrinsic ISRF of NGC4816,
but does not bias our conclusions (cf. Sect. 4.1).
Finally, a region of extremely low $\rm H \alpha$ line emission
is present at a radial distance of about 20.6 arcsec, on both sides
of the galaxy.
We tentatively interpret the latter region of mild star formation activity
as the signature of two symmetric spiral arms.

Fig. 7 details the morphology of the $\rm H \alpha$ line emission
across a strip of about $\rm 2.5 \times 50~arcsec^2$ centered on
the galaxy nucleus, but does not solve the question about the topology
of the sources of the $\rm H \alpha$ line emission associated with
the dust lane of NGC4826.
The interpretation of the CO map of this galaxy by Sakamoto et al.
(1999 - cf. Sect. 1) may give a hint that the ongoing star formation
activity across the Evil Eye is associated with the disk and obscured
by the molecular gas in the dark lane which has not yet settled in the disk.
Since this interpretation is questionable (cf. Sect. 1), we assume
two limiting cases for the topology of the sources of the asymmetric
$\rm H \alpha$ line emission.
In the first case (case ``a''), the latter sources are assumed to be located
below the dust lane, so that the excess (asymmetric) $\rm H \alpha$ line
intensity, as seen by the observer, must be corrected for absorption
and scattering by the dust in the Evil Eye, as well as the axi-symmetric
component of the $\rm H \alpha$ line emission, associated with the disk. 
Therefore, the total (axi-symmetric$+$asymmetric) $\rm H \alpha$ line flux,
$F(H \alpha)_{TOT}$, corrected for absorption and scattering by dust,
as seen by the observer is given by:
\[\rm F(H \alpha)_{TOT}^a = F(H \alpha)_{TOT}^{obs}~e^{\tau_{H \alpha}(att)},\]
where $F(H \alpha)_{TOT}^{obs}$ is the observed total $\rm H \alpha$ line
flux across the dust lane.
In the second case (case ``b''), we assume that the sources of the asymmetric
$\rm H \alpha$ line emission are on top of the dust lane,
so that their contribution to $F(H \alpha)_{TOT}$ is not corrected for
absorption and scattering by dust.
This time $F(H \alpha)_{TOT}$ is given by: 
\[\rm F(H \alpha)_{TOT}^b = F(H \alpha)_{TOT}^{obs} + F(H \alpha)_{s} (1 - e^{- \tau_{H \alpha}(att)}),\]
where $F(H \alpha)_{s}$ is the axi-symmetric component
of the $\rm H \alpha$ line emission.
The $\rm H \alpha$ line flux seen by the dust lane
is intermediate with respect to these two estimates.

The integrated (shortward of $\rm \le 912~\AA$) Lyman continuum (Lyc)
photon flux of a given OB-association is estimated from
the $\rm H \alpha$ line flux, by adopting a Lyc/$\rm H \alpha$ photon
conversion factor of 2.2 for case-B recombination (Osterbrock 1989)
with an assumed electron temperature $\rm T_e = 10^4~K$
(Hummer \& Storey 1987; Oey \& Kennicutt 1997).

We plot the radial variation of the total integrated Lyc photon flux
$Q(H^0)_{TOT}$ in Fig. 8.
Here the long- and short-dashed lines represent cases ``a'' and ``b'',
respectively, while the solid lines correspond to reference values
of the integrated Lyc photon flux (at the stellar surface) for representative
individual stars of spectral type O3 and O9.5 and of luminosity class
I, III and V.
These reference stellar values are derived by Schaerer \& de Koter (1997)
from their solar metallicity {\it CoStar} models, using the $T_{eff}$-log $g$
calibration of Vacca et al. (1996) and adopting the stellar radii
from Vacca et al..
The values of $Q(H^0)_{TOT}$ estimated for the Evil Eye fall nicely between
those of individual O3 -- B0 stars (cf. Schaerer \& de Koter),
and suggest that the Evil Eye probably does not host O-stars as UV-bright as
type O5I or brighter (i.e. with $Q(H^0)_{TOT} \rm \ge \sim 10^{50}~s^{-1}$),
though luminosities are uncertain by a factor of 4 (cf. Sect. 1).
Of course, the number of OB-stars of given spectral type (and with given
metallicity, effective temperature $T_{eff}$ and surface gravity $g$)
actually present in the different regions of the galaxy under study
is not constrained.
The radial variation of $Q(H^0)_{TOT}$ implies that the total number
of stars in the various OB-associations across the dark lane changes,
this change being significantly different between cases ``a'' and ``b''.

A better characterization of the hardness of the local radiation field
across the dust lane is given by the ratio of the [NII]
($\rm \lambda = 6583~\AA$) and the $\rm H \alpha$ line intensities,
$[NII](\lambda 6583)/H \alpha$, and of the intensity ratio between the [SII]
($\rm \lambda \lambda = 6716$, $\rm 6731~\AA$) doublet
and the $\rm H \alpha$ line, $[SII](\lambda \lambda 6716+6731)/H \alpha$.
In fact, these ratios depend also on the characteristic $T_{eff}$
of the OB-stars (e.g. Kennicutt et al. 2000).
We plot the radial behaviour of $[NII](\lambda 6583)/H \alpha$ 
and of $[SII](\lambda \lambda 6716+6731)/H \alpha$,
for cases ``a'' and ``b'', in Fig. 9.
Larger uncertainties affect the two ratios for $r =$ 1.3 arcsec,
due to the uncertain continuum baselines and deblending there,
and for radial distances larger than about 20.6 arcsec, where the differences
between the [NII]($\rm \lambda 6583$), [SII]($\rm \lambda \lambda 6716+6731$)
and $\rm H \alpha$ line intensities of the obscured and unobscured regions
across the galaxy nucleus are very small and of the same order of magnitude
of the observational errors.
This explains the sudden fall of $[SII](\lambda \lambda 6716+6731)/H \alpha$
for $r = 23.9$ arcsec in Fig. 9c.
Reassuringly, both these spectroscopic indices show the same overall
radial dependence, irrespective of the assumed location of the sources
of the extra $\rm H \alpha$ line emission with respect to the dust lane.
Changes of $[SII](\lambda \lambda 6716+6731)/H \alpha$
happen in a smoother way, probably because of the complementarity
of the [NII]($\rm \lambda 6583$) and $\rm H \alpha$ line fluxes,
due to the partial blending of these two lines at our spectral resolution.

Fig. 9b shows that $[NII](\lambda 6583)/ H \alpha$ drops from 0.2 dex
to $-0.8$ dex, moving from the nucleus to the neighborhood of the broad,
bright HII region, while it increases again for galactocentric distances
larger than about 18.1 arcsec, external to this HII region.
For the HII regions in our Galaxy and in the Magellanic Clouds,
values of $[NII](\lambda 6583)/H \alpha$ of about $-0.8$ dex
are characteristic of stellar effective temperatures between
35 and 45$\rm \times 10^3~K$, while
$-0.6 \le [NII](\lambda 6583)/H \alpha \le \rm -0.3~dex$ typically corresponds
to $T_{eff} \rm < 35 \times 10^3~K$ (Kennicutt et al. 2000).
Therefore the presence of solar metallicity stars of type O5 -- O8 seems 
to be allowed in the broad, luminous HII region associated with the Evil Eye,
while only stars later than O9 are probably present at other locations across
the dust lane (cf. Schaerer \& de Koter 1997, their Tab. 3) and, therefore,
are the sources of the H-ionization there.
Whatever the exact stellar classification of the OB-associations
in the obscured region of NGC4826 is, Fig. 9 confirms that the radial
increase of $Q(H^0)_{TOT}$ toward the broad, bright HII region in the Evil Eye
is associated with the hardening of the stellar UV spectrum.

The large values of $[NII](\lambda 6583)/H \alpha$
for $r \le 2.1$ arcsec are in excellent agreement with the range quoted
by Keel (1983b) ($-0.08 \le [NII](\lambda 6583)/H \alpha \le \rm 0.10~dex$)
for the nuclear region of NGC4826.
They exceed the upper values of $[NII](\lambda 6583)/H \alpha$
found in HII regions of our Galaxy and of the Magellanic Clouds
(Kennicutt et al. 2000) and, as concluded by Keel, they may point to
different sources of ionization and/or of the ionizing spectrum
in the NGC4826 nucleus, as provided by either an AGN or by
a starburst-powered LINER (cf. Ho et al. 1997; Alonso-Herrero et al. 2000).
By contrast, Keel quotes values of $[SII](\lambda \lambda 6716+6731)/H \alpha$
in the $-0.19$ -- $-0.11$ dex range, against our mean value of
$\sim 0.15$ dex within a distance of 2.35 arcsec from the NGC4826 nucleus.
This discrepancy may point to inaccurate line modeling and/or
subtraction of the continuum for the [SII]
($\rm \lambda \lambda = 6716+6731~\AA$) lines of our sub-spectra
for $r \le 2.1$ arcsec.
The large values of $[NII](\lambda 6583)/H \alpha$
and $[SII](\lambda \lambda 6716+6731)/H \alpha$ for $r \ge 22.3$ arcsec
do not find any counterparts in those reproduced in Kennicutt et al. (2000).
We suspect that these values of ours are biased by errors in deblending
and continuum subtraction, these lines having a low signal-to-noise ratio,
in addition.

Finally, in Fig. 10, we plot the radial distribution (across the dust lane)
of $I(ERE)$ (panel a) and of $I(ERE)/I(sca)$ (panel b), as derived
in Sect. 4.2, together with that of $Q(H^0)_{TOT}$ (panel c), for the two
limiting cases considered here.
The comparison of this plot with that in Fig. 5 is subject of interpretation
in Sect. 5.

\subsection{The continuum emission from the broad HII region in the Evil Eye}

The detailed study of the observed stellar SED of NGC4826 involves
the comparison with model stellar SEDs and is beyond the goal
of the present investigation.
Nonetheless, the (admittedly rough) information about the spectral types
of the stars responsible of the H-ionization (cf. Sect. 4.4) provides
an order-of-magnitude estimate of the contribution of these stars to
the residual blue emission found in the sub-spectra of the broad
HII region of the Evil Eye intercepted by the spectrograph slit
(cf. Sect. 4.1).
For simplicity, here we assume that the H-ionizing stars present there form
a simple stellar population with solar metallicity.

As an initial working hypothesis, we assume that all these stars may be
represented by only one spectral type, identified by the average
$T_{eff}$ of 40000 K (cf. Sect. 4.4).
Such a star corresponds to a 2.23 Myrs old star with initial mass
of 60 $\rm M_{\odot}$, classified as spectral type O5.5IV (on the basis
of its location in the HR-diagram), according to the {\it CoStar} model D3
(Schaerer \& de Koter 1997).
The Lyc photon flux of this model star may be easily derived from the data
listed by Schaerer \& de Koter.
As a result, we find that one model D3 star may provide 60\% of the total
Lyc photon flux produced within the HII region at e.g. $r = 13.9$ arcsec
in case ``a'' (cf. Sect. 4.4), equal to $\rm 1.06 \times 10^{50}~s^{-1}$.
By contrast, only O-stars less UV-bright than one model D3 star may provide
the total Lyc photon flux produced there in case ``b'' (cf. Sect. 4.4),
equal to $\rm 2.57 \times 10^{49}~s^{-1}$.
A more complex representation of the OB-associations across the Evil Eye
is needed.
Therefore, now we assume that these OB-associations are represented by
two ensembles, one associated with $T_{eff}$ within 35
and 45$\rm \times 10^3~K$ and the other with $T_{eff} \rm < 35\times 10^3~K$,
consistent with our data (cf. Sect. 4.4).
Schaerer \& de Koter provide only two models which satisfy all
our requirements: one (model B2) corresponds to a 2.60 Myrs old star
with initial mass of 25 $\rm M_{\odot}$ and $T_{eff} \rm = 36300~K$,
classified as spectral type O9V; the other (model D4) corresponds to
a 2.76 Myrs old star with initial mass of 60 $\rm M_{\odot}$
and $T_{eff} \rm = 32200~K$, classified as spectral type O7I.
For our illustrative purposes this is enough.
We note that the number of model B2 objects should be at least 3 times
larger than that of model D4 objects in order to satisfy our observational
constraint on the effective temperatures of the H-ionizing stars within
the broad HII region of NGC4826 (Fig. 9b).
On the other hand, the number of 25 $\rm M_{\odot}$ (model B2) stars
should be about 8 times larger than that of 60 $\rm M_{\odot}$ (model D4)
stars in case of e.g. a Salpeter (1955) initial mass function (IMF).
Model D4 has a Lyc photon flux about 7 times larger than model B2
(Schaerer \& de Koter), so that, for a Salpeter IMF, we may expect that
equipartition of the total Lyc photon flux between these two ensembles
applies, to zeroth order.
Due to the discrete nature of stars, we find that either a mix of 24 model B2
stars and 2 model D4 stars or a mix of 17 model B2 stars and 3 model D4 stars
are reasonable solutions for case ``a''.
By contrast, either 2 model B2 stars plus one model D4 star
or 10 model B2 stars alone may provide the needed total Lyc photon flux
for case ``b''.
In both cases, we can predict the observed SEDs of the OB-association
within the broad HII region of NGC4826 at $r = 13.9$ arcsec from
the knowledge of the optical depth at different wavelengths there
(cf. Sect. 3.2), given the intrinsic SEDs of the model stars
(D. Schaerer 2001, private communication).
In Fig. 11, we show the predicted profile of the total continuum emission
for case ``a'' (with extinction applied) and that for case ``b'',
together with the spectrum of the residuals already displayed in Fig. 4h,
here reproduced by using a magnified intensity scale.
The predicted total continuum emission is the sum of the atomic continuum
emission from the gas (derived hereafter) and the continuum emission arising
either from the mix of 17 model B2 stars and 3 model D4 stars, for case ``a'',
or from the mix of 2 model B2 stars and 1 model D4 star, for case ``b''.
For case ``a'' the predicted total continuum emission is an upper limit,
as we will explain.

The comparison between model predictions and data shows that, in either case,
{\it up to} about 50\% of the residual blue emission can be accounted for
by the emission of the two ensembles of H-ionizing stars alone.
At $\rm \lambda \sim 7500~\AA$, the predicted stellar emission is of
the same order of magnitude of the rms uncertainty of the background
subtracted in the spectrum extraction procedure (Sect. 3.1).
It is also {\it at least} of the same order of magnitude of
the estimated observed atomic continuum emission, as we demonstrate hereafter.

The nebular continuum intensity $I_0(\lambda)$ relative to the $\rm H \beta$
($\rm \lambda = 4861~\AA$) line intensity $I_0(H \beta)$ corrected for
interstellar extinction, may be espressed as a function of the observed
nebular continuum intensity $I(\lambda)$ and observed $\rm H \beta$ line
intensity $I(H \beta)$ according to the classical formula:
\[\rm I_0(\lambda)/I_0(H \beta) = I(\lambda)/I(H \beta) \times 10^{C(H \beta) f(\lambda)},\]
where $C(H \beta)$ and $f(\lambda)$ are the extinction correction
at $\rm H \beta$ and the extinction correction function, respectively.
In the 5300 -- $\rm 9100~\AA$ spectral range, the nebular continuum intensity
profile may be approximated by two top-hat functions, the intensity
shortward of $\rm \sim 8200~\AA$ being three times higher than longward 
of $\rm \sim 8200~\AA$ (cf. Sect. 3.2).
We set $I_0(\lambda)/I_0(H \beta)$ equal to 0.015
in the 5300 -- $\rm 8200~\AA$ spectral range, this ratio being found
by Sivan \& Perrin (1993) for the ``Bubble Nebula'' (NGC7635).
Since our spectral range does not cover the region of the $\rm H \beta$ line,
we derive $I_0(H \beta)$ and $I(H \beta)$ from the analogous values of
the $\rm H \alpha$ line intensity.
At $r = 13.9$ arcsec, the observed $\rm H \alpha$ line intensity
associated with the broad HII region in the Evil Eye is equal to
$\rm \sim 1.23 \times 10^{-5}~ergs~s^{-1}~\AA^{-1}~cm^{-2}~sr^{-1}$.
If the nebular emission arises from the top of the dust lane (case ``b''),
it is not affected by extinction and, therefore,
$I(\lambda)/I(H \beta)=I_0(\lambda)/I_0(H \beta)=0.015$
in the 5300 -- $\rm 8200~\AA$ spectral range,
and the $\rm H \beta$-to-$\rm H \alpha$ line ratio is equal to its
theoretical value.
For electron densities $n_e \rm \le 10^4~cm^{-3}$ and electron temperatures
$T_e \rm \sim 10^4~K$, the theoretical $\rm H \beta$-to-$\rm H \alpha$ line
ratio is equal to 0.351 (Osterbrock 1989).
As a consequence, the predicted nebular continuum intensity (as seen by
the observer) is
$\rm \sim 6.5 \times 10^{-8}~ergs~s^{-1}~\AA^{-1}~cm^{-2}~sr^{-1}$
in the 5300 -- $\rm 8200~\AA$ spectral range
and $\rm \sim 2.2 \times 10^{-8}~ergs~s^{-1}~\AA^{-1}~cm^{-2}~sr^{-1}$
in the $\sim 8200$ -- $\rm 9100~\AA$ spectral range, for case ``b''.
It is straightforward to see that, when extinction matters (case ``a''),
$I(\lambda)/I(H \beta) < 0.015$ in the 5300 -- $\rm 8200~\AA$ spectral range
and the observed $\rm H \beta$-to-$\rm H \alpha$ line ratio is less
than 0.351.
Therefore, the predicted nebular continuum intensity (as seen by
the observer) is less than 
$\rm \sim 6.5 \times 10^{-8}~ergs~s^{-1}~\AA^{-1}~cm^{-2}~sr^{-1}$
in the 5300 -- $\rm 8200~\AA$ spectral range and less than
$\rm \sim 2.2 \times 10^{-8}~ergs~s^{-1}~\AA^{-1}~cm^{-2}~sr^{-1}$
in the $\sim 8200$ -- $\rm 9100~\AA$ spectral range, for case ``a''.
As anticipated in Sect. 3.2, in the broad HII region of the Evil Eye
under investigation, the observed nebular continuum emission is expected
to be {\it at maximum} equal to the observed emission from the H-ionizing
stars {\it only} at wavelengths larger than $\rm \sim 7500~\AA$.

From the consideration of the poor assumptions made here and the significant
results obtained from them, we conclude that a mixture of H-ionizing
stars with different spectral types, in reasonable number (considering
the region of $\rm 20.4 \times 60.6~pc^2$ area sampled per pixel),
is responsible of most of the residual blue emission found
in the sub-spectra of this HII region.
Reassuringly, the total continuum emission from H-ionizing stars and the gas
in the 7800 -- 9100 $\rm \AA$ spectral region, where the fit in Fig. 4h
is made, is at maximum of the same order of magnitude of the rms uncertainty
of the background subtracted in the spectrum extraction procedure (Sect. 3.1)
and has a flat spectrum between 8200 and 9100 $\rm \AA$.
As a consequence, we conclude that optical depths may be systematically,
but only slightly, underestimated when traversing the broad HII region
in the Evil Eye intercepted by the spectrograph slit, but that the detection
of ERE there is not compromised by the neglect of the total continuum emission
from H-ionizing stars and the gas.

\section{Discussion}

The detection of ERE from the dust lane of NGC4826 has been validated
from the analysis reported in Sect. 3.2, 4.1 and 4.5.
In addition to our own Galaxy, M82 and LMC (cf. Sect. 1), the Evil Eye Galaxy
is the fourth galaxy known so far to display ERE.
We note that different methods were adopted for the detection of ERE
in these galaxies.
Since the spatial information about individual sources of ERE in our Galaxy
is still un-matched by observations of external galaxies, it may make
more sense to compare the ERE phenomenology of NGC4826
with those of the halo of M82 and 30-Doradus in the LMC.
This may be also suggested by the starburst nature of these three galaxies
or, at least, of their observed regions (cf. Sect. 1).
However, the more extensive ERE phenomenology displayed by the Milky Way
is a good basis to the understanding of the ERE phenomenon and is taken
into account hereafter.

In the halo of M82, ERE is found to be centered at about $\rm 6900~\AA$,
with a FWHM of approximately $\rm 1500~\AA$ (Perrin, Darbon \& Sivan 1995).
In 30-Doradus, the ERE band peaks around $\rm 7270~\AA$,
is $\rm 1140~\AA$ wide (Darbon, Perrin \& Sivan 1998) and also looks
similar to those found in Galactic HII regions (Perrin \& Sivan 1992;
Sivan \& Perrin 1993; Darbon et al. 2000).
In the Evil Eye, we were able to determine radial changes of $\lambda_p(ERE)$,
as well as of the wavelength range of the ERE (Sect. 4.2) for the first time
in an external galaxy.
In particular, $\lambda_p(ERE) \rm \sim 8300~\AA$ in the innermost region
of the Evil Eye, and shifts longward of $\rm 9100~\AA$,
at galactocentric distances larger than 8.8 arcsec.
A larger value of $\lambda_p(ERE)$ is associated with a larger characteristic
size of the photoluminescing particles, as found by Ledoux et al. (2000)
for silicon nanocrystals.
Under the assumption that the ERE carriers are identified
with silicon nanoparticles (e.g. Witt, Gordon \& Furton 1998),
the results of Ledoux et al. lead us to the conclusion that
the active ERE carriers in the Evil Eye have a characteristic size
of about 5 nm or larger, i.e. about 40\% larger than those
in the halo of M82 and almost two times larger than those
in the Galactic diffuse ISM (cf. Szomoru \& Guhathakurta 1998).
Unfortunately, the spectral range of our observations does not allow us
to recover the whole ERE band-width.
We estimate an extraordinarily large FWHM of approximately $\rm 2000~\AA$
from fitting the observed SED of the ERE in the innermost region of
the Evil Eye with a symmetry argument, whose validity may be questionable.
However, for the Evil Eye, the finding of large values both of the ERE band
FWHM and of $\lambda_p(ERE)$ is consistent with the existence of a relation
between these two quantities (Darbon, Perrin \& Sivan 1999).
For reference, in the Orion Nebula, ERE peaks around $\rm 8200~\AA$,
with a $\rm 1800~\AA$ FWHM (Perrin \& Sivan 1992).
We also find a maximum value of $I_p(ERE)$ of about
$\rm 3.7 \times 10^{-6}~ergs~s^{-1}~\AA^{-1}~cm^{-2}~sr^{-1}$,
for a galactocentric distance $r = 1.3$ arcsec, about four times higher
than that derived in the halo of M82
($\rm \sim 1 \times 10^{-6}~ergs~s^{-1}~\AA^{-1}~cm^{-2}~sr^{-1}$).
In the Orion Nebula, $I_p(ERE)$ is about
$\rm 3.5 \times 10^{-6}~ergs~s^{-1}~\AA^{-1}~cm^{-2}~sr^{-1}$
(Perrin \& Sivan 1992), i.e. almost the same as the maximum value
of $I_p(ERE)$ in the dark lane of NGC4826, with $I(ERE)/I(sca)$ of about 0.6
(Gordon, Witt \& Friedmann 1998).
In the Evil Eye, $I_p(ERE)$ decreases to a value of
$\rm \sim 2.3 \times 10^{-6}~ergs~s^{-1}~\AA^{-1}~cm^{-2}~sr^{-1}$,
for $r = 4.6$ arcsec, and keeps on decreasing farther out.
At larger galactocentric distances, where there is a broad, bright HII region
associated with the dust lane (Sect. 4.2), ERE, if any, has to be faint
and beyond the spectral range of our observations.
No ERE is detected at even larger galactocentric distances,
within the spectral range and noise threshold of our observations.

Since $I_p(ERE)$ is to first order a measure of the illuminating
UV/optical radiation field intensity, as in any other photoluminescence
phenomenon, we conclude that the strength of the UV/optical light
in the innermost region of the Evil Eye is about four times larger than
in the halo of M82 and almost the same as in the Orion Nebula.
However, from the comparison of Fig. 5 and Fig. 10, it emerges that
the ERE intensity is not simply given by the product of the illuminating
UV/optical radiation field times the dust column density,
as it might be expected.
In fact, both $I(ERE)$ and $I(ERE)/I(sca)$ do not show a monotonic behaviour
with increasing values of $\tau_V(att)$ (Fig. 5) and of $Q(H^0)_{TOT}$
(Fig. 10).
We remind the reader that $I(ERE)$ is proportional to the radiation field
shortward of $\rm 5500~\AA$ and $I(sca)$ is proportional to the number
of optical photons longward of $\rm 5500~\AA$.
Thus the bluer the illuminating radiation field, the larger $I(ERE)/I(sca)$.
Conversely, $I(ERE)$ and $I(ERE)/I(sca)$ tend to 0, at least within
the spectral range and noise threshold of our observations, as a broad,
bright HII region is approached.
The interpretation of the complex behaviour of the ERE intensity with
optical depth and UV/optical radiation field intensity has been recently
investigated theoretically by Smith \& Witt (2001).
These authors show how the experimentally established recognition
that photoionization quenches the luminescence of silicon nanoparticles
(SNPs) may play a key-role in the reproduction of the extensive phenomenology
of ERE in different astrophysical environments of our Galaxy.
Thus, it is the excitation and ionization equilibrium of SNPs under the
interstellar conditions of a given environment that regulates the fraction
of neutral ERE carriers, the only ones capable of luminescing, there.
According to this model, the shift of the ERE band's peak wavelength
toward larger values with increasing UV/optical radiation density requires
a change of the size distribution of the actively luminescing ERE carriers.
This is accomplished by the selective removal of the smaller particles
due to size-dependent photofragmentation.
We refer the reader to the work of Smith \& Witt in order to understand
the photophysics of the ERE carriers in more detail.
Hereafter we interpret our observational results through this model.

First, it is straightforward to conclude that the presence of a broad,
bright HII region, in the portion of the dust lane of NGC4826 subject
to our spectroscopic observation, has one main consequence
on the ERE phenomenology, illustrated in Fig. 4a--n: no ERE is detected
(within our spectral range and noise threshold) between the inner edge
of this HII region and the outer edge of the dust lane (Fig. 4h--n).
Since the optical depth is relatively large in the vicinity of
the HII region (Fig. 5), we conclude that the observed absence of ERE there
is not a result of a lack in dust but may be consistent with
the photoionization-induced quenching of the luminescence in silicon
nanoparticles and the effect of photofragmentation on the characteristic size
of the (neutral) luminescing particles.
Indeed a hard and strong UV radiation field is present there, as witnessed
by the hot effective temperatures and the ionizing photon rate
of the OB-associations (Sect. 4.4).

This conclusion seems at odds with the detection of ERE from some of
the observed regions of the interstellar medium within Galactic HII regions
like Orion (Perrin \& Sivan 1992), the Bubble Nebula (Sivan \& Perrin 1993)
and Sh152 (Darbon et al. 2000) or within 30-Doradus in the LMC
(Darbon, Perrin \& Sivan 1998).
In particular, ERE with peak intensity of about
4 -- 6 $\rm \times 10^{-7}~ergs~s^{-1}~\AA^{-1}~cm^{-2}~sr^{-1}$,
peak wavelength around $\rm 7200~\AA$ and band-width around $\rm 1000~\AA$
has been found to be spatially associated with the ionized gas in Sh152
(Darbon et al. 2000).
Nonetheless, in this HII region, no significant ERE is detected where
the $\rm H \alpha$-line surface brightness peaks, i.e. at the location
of the ionizing O9V star S152.1 (Fig. 7 in Darbon et al. 2000).
Under the assumption that ERE with
$I_p(ERE) \sim \rm 5 \times 10^{-7}~ergs~s^{-1}~\AA^{-1}~cm^{-2}~sr^{-1}$
around $\rm 7200~\AA$ arises from the HII region of NGC4826
under discussion, this peak intensity would result to be one order
of magnitude higher than the rms uncertainty of the background subtracted
around this wavelength (cf. Sect. 3.1) and, therefore, detectable.
Fig. 11 shows that the ratio of the residual intensity and the rms
uncertainty of the background subtracted at around $\rm 7200~\AA$
is only about two.
By contrast, ERE intensities of less than
$\rm 10^{-6}~ergs~s^{-1}~\AA^{-1}~cm^{-2}~sr^{-1}$ are detected by us
at $\rm \lambda \ge \sim 7500~\AA$ for $r = 10.5$ arcsec (Fig. 4f),
and ERE intensities of less than
$\rm \sim 5 \times 10^{-7}~ergs~s^{-1}~\AA^{-1}~cm^{-2}~sr^{-1}$
are tentatively detected at $\rm \lambda \ge \sim 8500~\AA$
for $r = 12.2$ arcsec (Fig. 4g).
Therefore, we conclude that, whether there is ERE across this HII region,
it has to be longward of $\rm 9100~\AA$.
A priori there is no reason to believe that the astrophysical conditions
in Orion, the Bubble Nebula, Sh152 and 30-Doradus are the same as
in the HII region of the Evil Eye under discussion and even homogeneous
across each of them.
Indeed the ERE phenomenology is different even for these four HII regions
where ERE was detected.

If there is no detection of ERE in the broad, bright HII region
intercepted by the spectrograph slit, strong ERE (i.e. with
$I_p(ERE) \rm \sim 3.7 \times 10^{-6}~ergs~s^{-1}~\AA^{-1}~cm^{-2}~sr^{-1}$)
is found near the nuclear region of NGC4826.
The high values of $I_p(ERE)$ and of $I(ERE)/I(sca)$ ($\sim 0.7$) measured
there may point either to a maximum fraction of neutral SNPs or
to a maximum ERE efficiency at this location.
The NGC4826 nucleus probably hosts different sources of H-ionization,
probably associated with a starburst-powered LINER, and has a minimum
dust column density.
Therefore a particularly high ERE efficiency associated with
a small population of neutral SNPs seems to be a viable explanation
of the ERE phenomenology there (cf. the case of Orion discussed
by Smith \& Witt).
Conversely, photoionization of the ERE carrier may justify the drop
of $I(ERE)$ and $I(ERE)/I(sca)$ for $r$ increasing from 1.3 to 4.6 arcsec,
where $Q(H^0)_{TOT}$ has a (probably secondary) peak (Fig. 10),
$\tau_V(att)$ increasing continuously there (Fig. 5).
A shallower decrease of $I(ERE)$ and an increase of $I(ERE)/I(sca)$
take place for 4.6 $< r \le$ 8.0 arcsec, where $Q(H^0)_{TOT}$ is approximately
constant and $\tau_V(att)$ still increases, with no shift of the ERE band.
This behaviour may reasonably be due to an increase of
the recombination rate, i.e. of the neutral fraction of the ensemble
of silicon nanoparticles, the only one capable of luminescing.
For $r >$ 8.0 arcsec, the once again increasing strength (probably)
and hardness of the UV radiation field cause the decline of both $I(ERE)$
and $I(ERE)/I(sca)$ and the shift of the ERE band toward longer wavelengths,
since the optical depth is about constant there.
At the inner edge of the HII region (for $r \sim$ 13 arcsec), $I(ERE)$
and $I(ERE)/I(sca)$ are equal to 0, as previously discussed.
The absence of ERE in the outermost regions of the Evil Eye,
external to this HII region, is consistent with the drop of the dust column
density and of the UV/optical radiation field (as suggested
by the precipitous drop of the integrated Lyc photon flux).

According to Smith \& Witt, the size-dependence of multiple ionization,
coupled with single-photon heating, may determine eventually
the characteristic size of the (neutral) luminescing particles.
In agreement with the results of the latter authors and with
the size-dependent photoluminescence properties of silicon nanocrystals
(Ledoux et al. 1998), in the Evil Eye both the ERE band and $\lambda_p(ERE)$
shift to longer wavelengths (cf. Fig. 4a--n) with increasing $r$,
approaching the HII region, as previously shown.
Indeed the values of $\lambda_p(ERE)$ of NGC4826 are even larger than
in the 30-Doradus, Bubble Nebula and Orion HII regions.
Aside from the exceptional size distribution of the actively luminescing
particles, the Evil Eye shows similar values of $I(ERE)/I(sca)$
(from 0.7 to 0, with increasing values of $r$) to those spanned
by different Galactic objects, like the diffuse ISM, reflection nebulae
and the Orion Nebula (cf. Gordon, Witt \& Friedmann 1998).
Relying on the results of the latter authors, we conclude that the ERE photon
conversion efficiency in NGC4826 is as high as found elsewhere.

\section{Conclusions}

NGC4826 is a nearby Sab galaxy with an outstanding asymmetrically placed,
absorbing dust lane (called the ``Evil Eye'') across its prominent bulge,
associated with several regions of ongoing star formation.
For this galaxy, we obtained accurate low-resolution ($\rm 4.3~\AA$/pixel)
long-slit spectroscopy (KPNO 4-Meter) in the 5300 -- $\rm 9100~\AA$
wavelength range, with a slit of 4.4 arcmin length, encompassing
the galaxy's bulge size, positioned across its nucleus.
This allowed us to study the wavelength-dependent effects of absorption
and scattering by the dust by comparing the stellar spectral energy
distributions of pairs of positions across the bulge, symmetrically placed
with respect to the nucleus, under the assumption that the intrinsic
(i.e. unobscured) radiation field is axi-symmetric, except for
the extra contribution due to ongoing star formation activity associated
with the dust lane.
In this analysis we made use of the multiple-scattering radiative
transfer calculations of Witt \& Gordon (2000).

As a main result, strong residual Extended Red Emission (ERE) was detected
from a region of the Evil Eye within about 13 arcsec projected distance
from the NGC4826 nucleus, adjacent to a broad, bright HII region,
intercepted by the spectrograph slit.

In the innermost portion of the dust lane, the ERE band extends
from about 5700 to $\rm 9100~\AA$, with an estimated peak
intensity near $\rm 8300~\AA$.
Here, the ERE-to-scattered light band-integrated intensity ratio
is about 0.7.

At farther galactocentric distances, approaching the broad, bright
HII region, the ERE band and peak intensity shift toward longer wavelengths,
while the ERE band-integrated intensity diminishes.
Finally, no ERE is detected within our spectral range and noise threshold
at the inner edge of this HII region and at any other position
at larger galactocentric distance.
A secondary maximum of the ERE-to-scattered light band-integrated
intensity ratio is found near the position of maximum opacity
across the dust lane, associated with a secondary maximum
of ongoing star formation activity.

These variations reveal a complex dependence of the ERE on
dust column density and interstellar radiation field.
While the former quantity is derived from modeling, our data do not
characterize the UV/optical flux shortward of $\rm 5300~\AA$. 
However, we are able to estimate the integrated (i.e. shortward
of $\rm 912~\AA$) Lyman continuum photon rate and the characteristic
effective temperatures of the OB-stars at its origin from the $\rm H \alpha$
($\rm \lambda = 6563~\AA$) line intensity and from the line intensity ratios
$[NII](\lambda~6583)/H \alpha$
and $[SII](\lambda \lambda~6716+6731)/H \alpha$.

We interpret the ERE as originating from photoluminescence by nanometer-sized
oxygen-passivated silicon particles, illuminated by UV/visible photons
of the local radiation field.
The phenomenology of the ERE in the Evil Eye is consistently interpreted
through the recent model of the photophysics of the ERE carrier
by Smith \& Witt (2001), which attributes a key-role to
the experimentally established recognition that photoionization quenches
the luminescence of silicon nanoparticles.

When examined within the context of ERE observations in the diffuse ISM
of our Galaxy and in a variety of other dusty environments such as
reflection nebulae and the Orion Nebula, we conclude that the ERE photon
conversion efficiency in NGC4826 is as high as found elsewhere, but that
the size of the actively luminescing nanoparticles in NGC4826 is about
twice as large as those thought to exist in the diffuse ISM of our Galaxy.

\acknowledgments
{\it Acknowledgments.} We are grateful to Rhodri Evans and Judith S. Young
for providing us with, respectively, the V-band image and the $\rm H \alpha$
image of NGC4826 here displayed.
We are also grateful to Daniel Schaerer for making available to us
the {\it CoStar} 1997 spectral energy distributions of OB-stars.
We acknowledge the contribution of Uma Vijh to the making of the pictorial
representation of the region of NGC4826 under investigation.
This work was supported through Grant NAG5-9202 from the National Aeronautics
and Space Administration to the University of Toledo.

\clearpage


\begin{figure}
\plotone{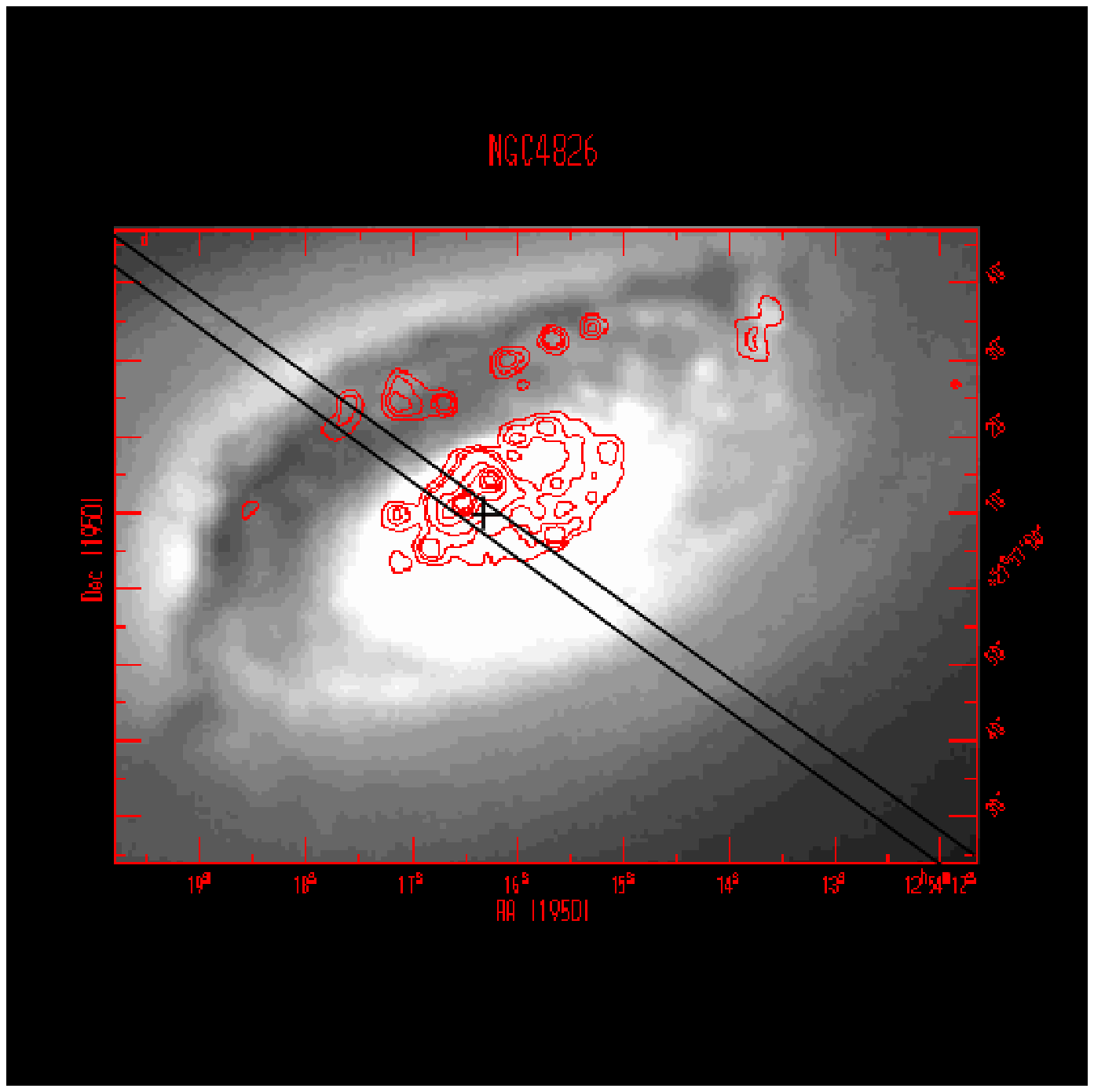}
\caption{Pictorial representation of the region of NGC4826 under
investigation, as intercepted by the spectrograph slit. Isophotal contours
of the galaxy emission in the $\rm H \alpha$ line (J. Young 2000,
private communication) are superposed onto a grey-scale two-dimensional
representation of the galaxy V-band surface brightness distribution
(R. Evans 2000, private communication). A cross marks the adopted position
of the galaxy center.\label{fig1}}
\end{figure}

\clearpage

\begin{figure}
\plotone{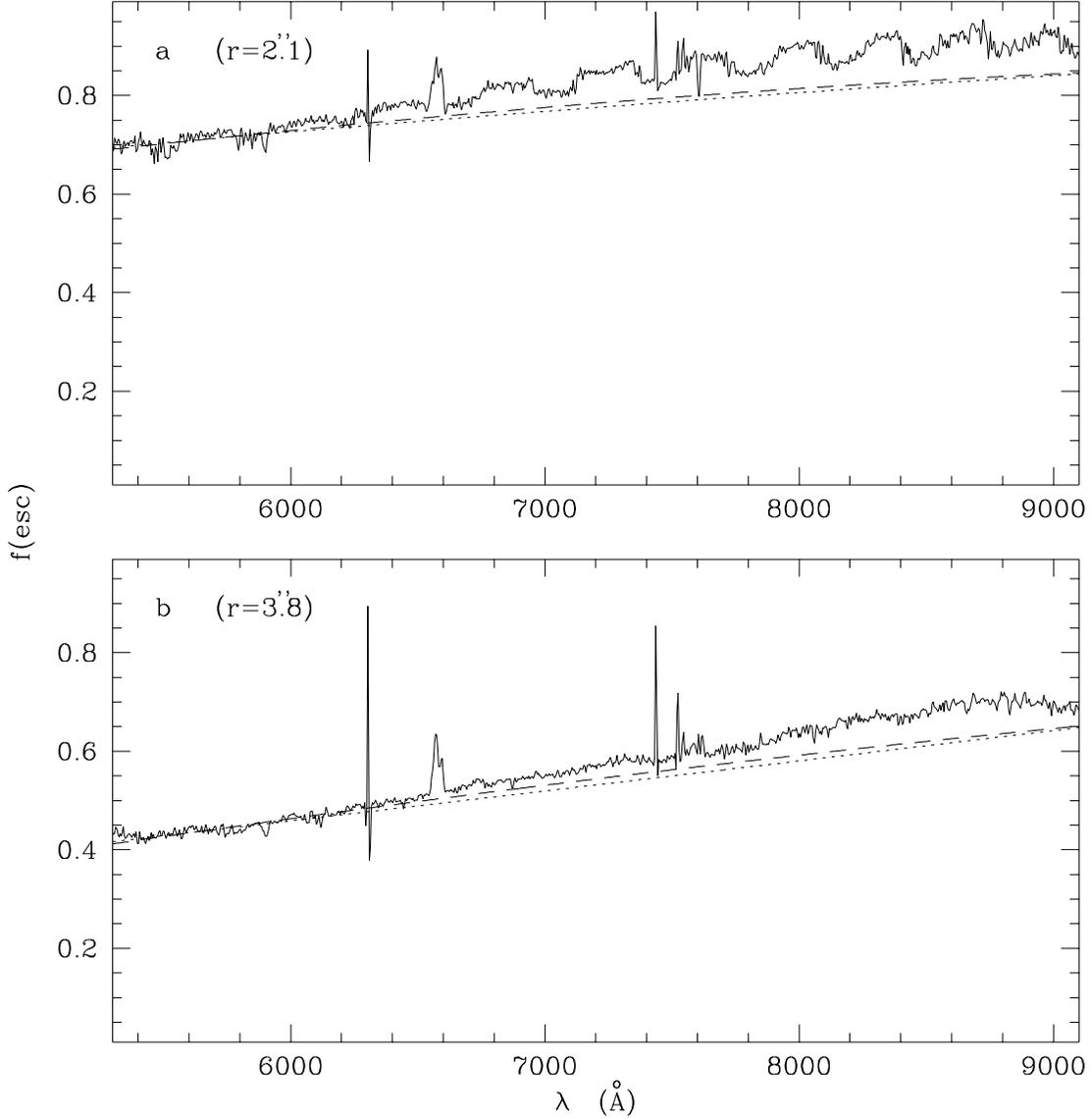}
\caption{The ratio of the observed spectral energy distributions
(SEDs) (solid line) for a pair of positions symmetrically located
at a projected distance $r$ with respect to the NGC4826 nucleus,
on the obscured and unobscured side, respectively.
Best-fit WG00 model curves of this ratio (see text), representing
$f_{\lambda}(esc)$, i.e. the fractional flux density distribution
of the light either directly transmitted or scattered by the dust,
are shown both for MW-type dust (dotted line) and for SMC-type dust
(short-dashed line).
From panel a through n, $r$ increases from 2.1 to 23.9 arcsec
in steps of 1.7 arcsec.\label{fig2}}
\end{figure}

\clearpage

\begin{figure}
\figurenum{2}
\plotone{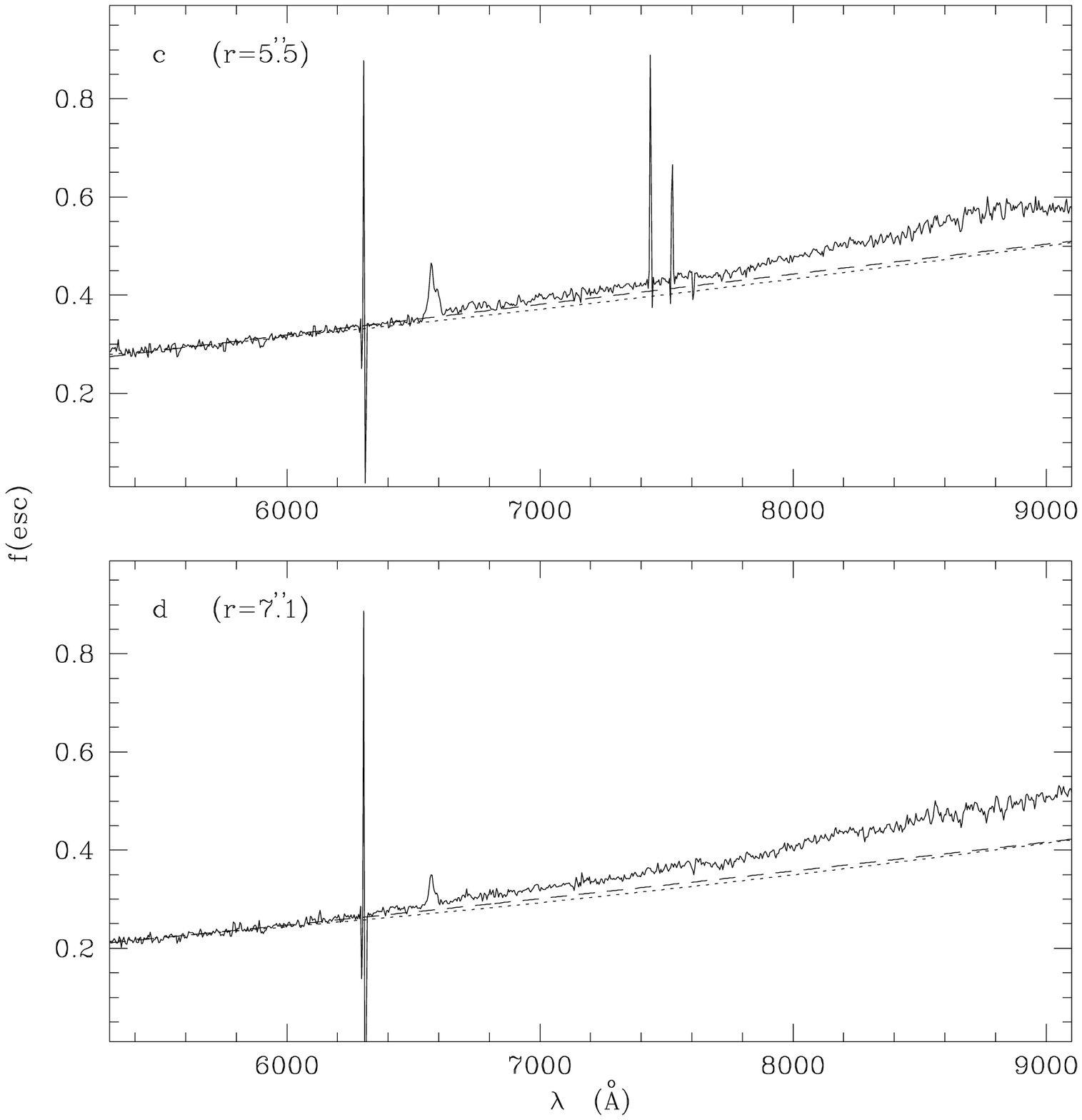}
\caption{continue}
\end{figure}

\clearpage

\begin{figure}
\figurenum{2}
\plotone{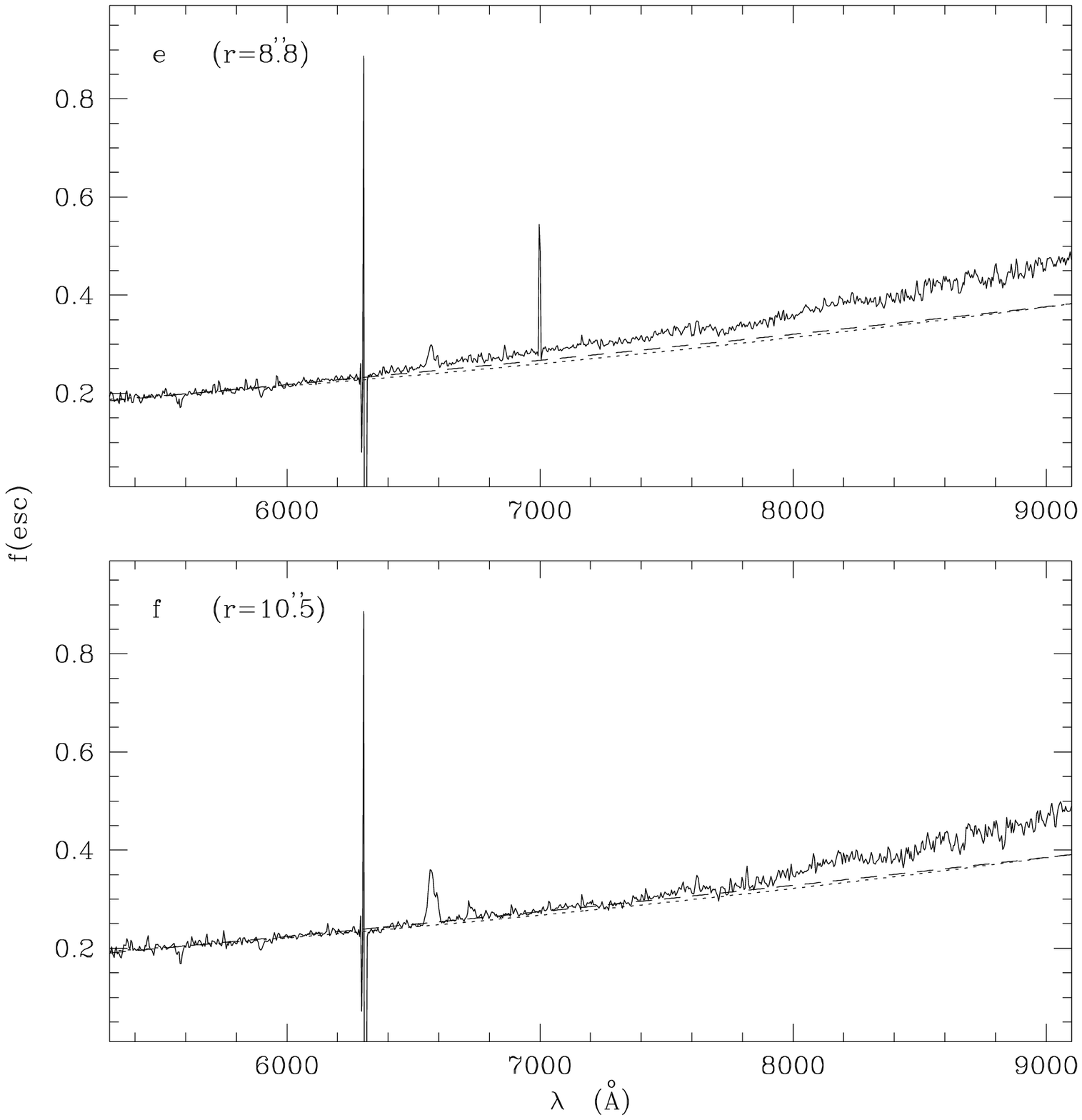}
\caption{continue}
\end{figure}

\clearpage

\begin{figure}
\figurenum{2}
\plotone{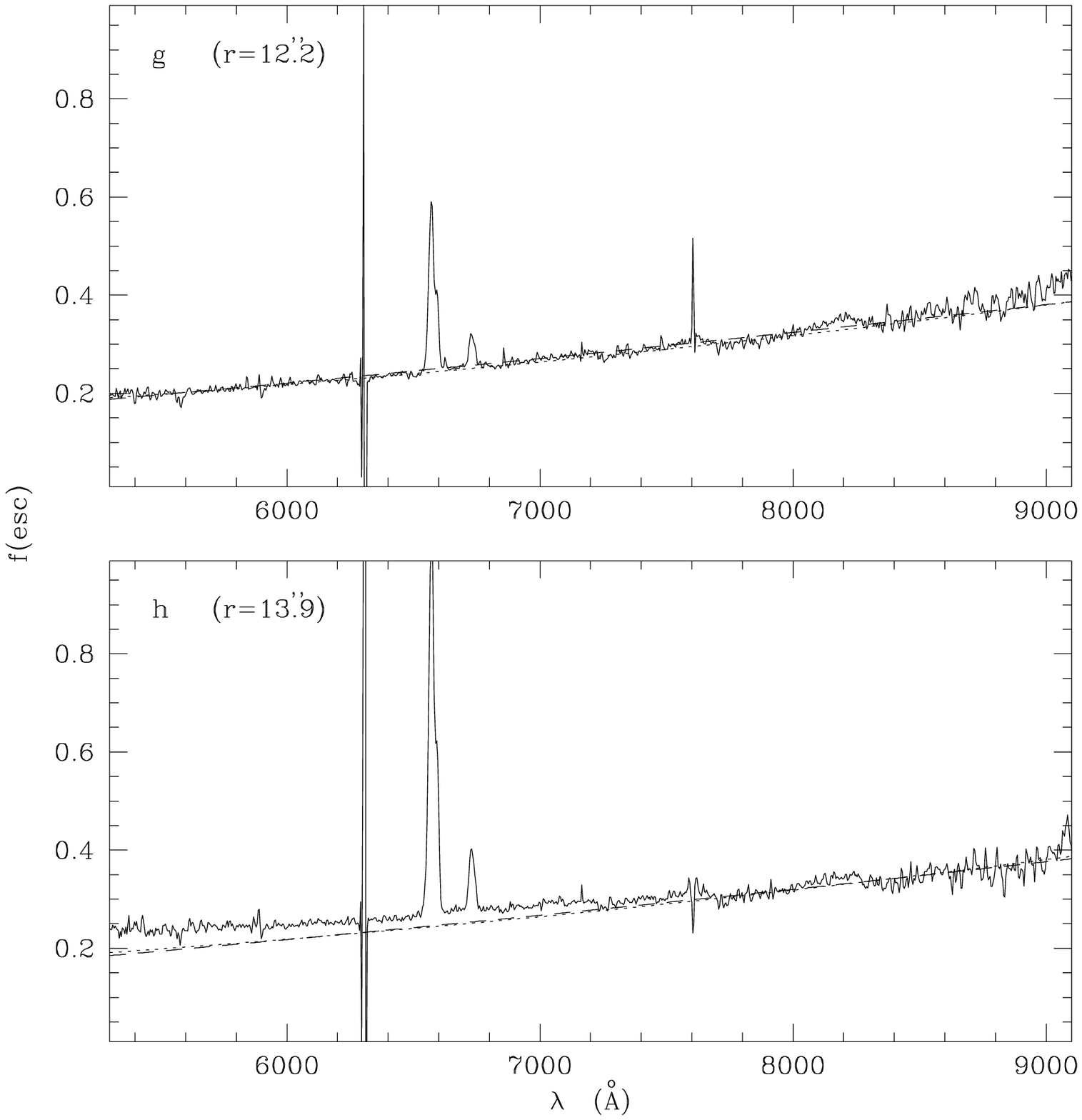}
\caption{continue}
\end{figure}

\clearpage

\begin{figure}
\figurenum{2}
\plotone{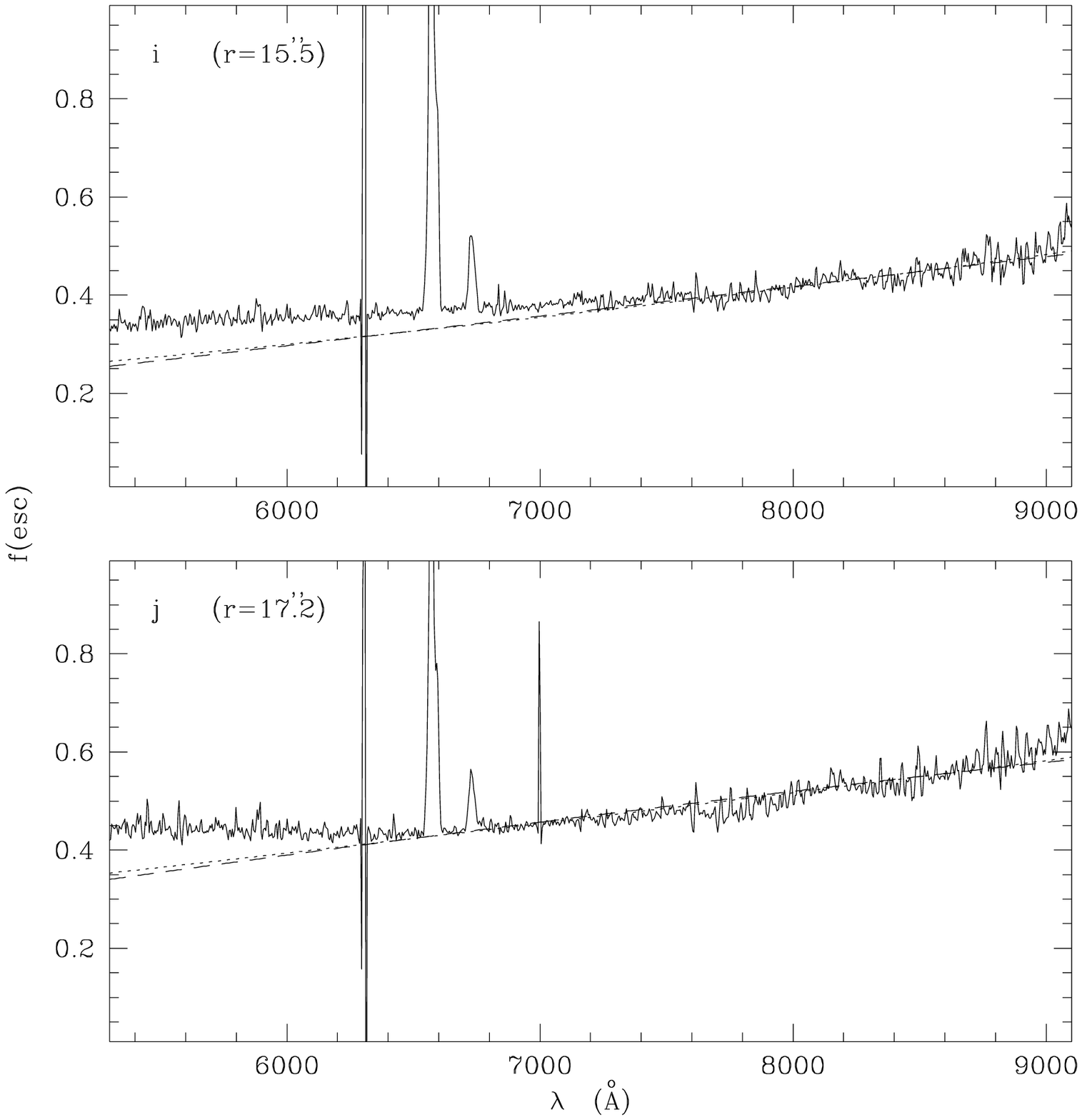}
\caption{continue}
\end{figure}

\clearpage

\begin{figure}
\figurenum{2}
\plotone{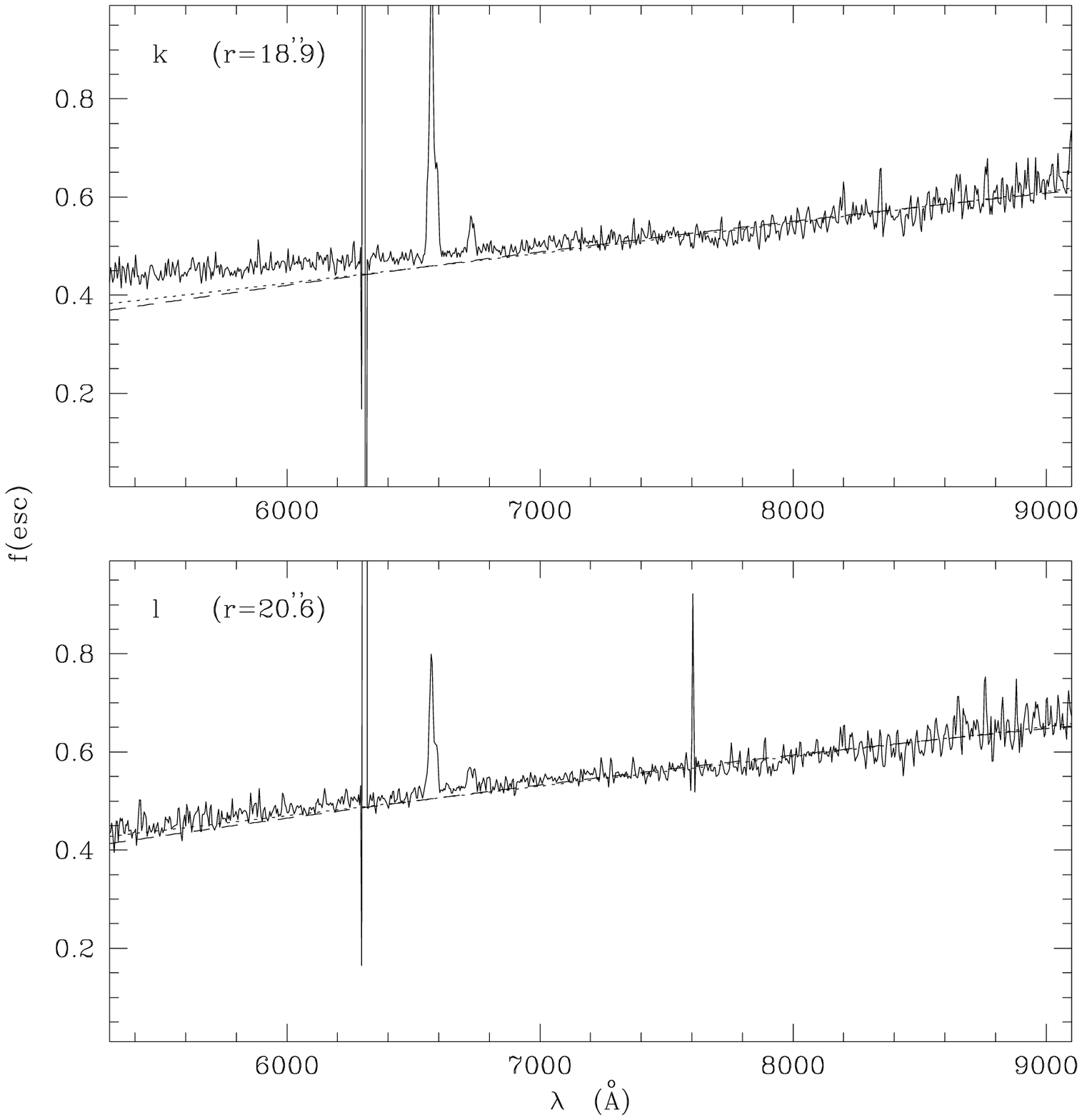}
\caption{continue}
\end{figure}

\clearpage

\begin{figure}
\figurenum{2}
\plotone{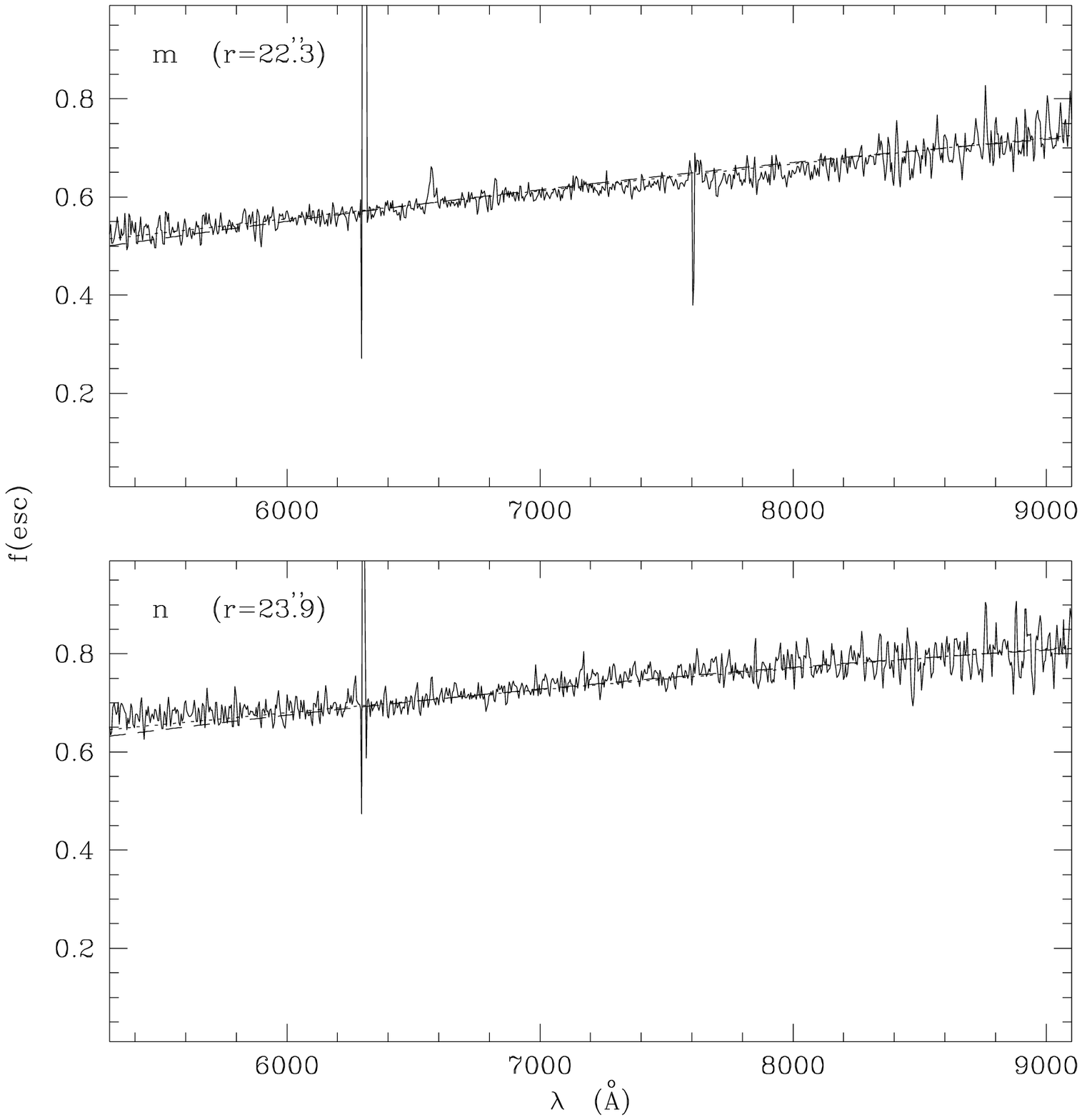}
\caption{continue}
\end{figure}

\clearpage

\begin{figure}
\plotone{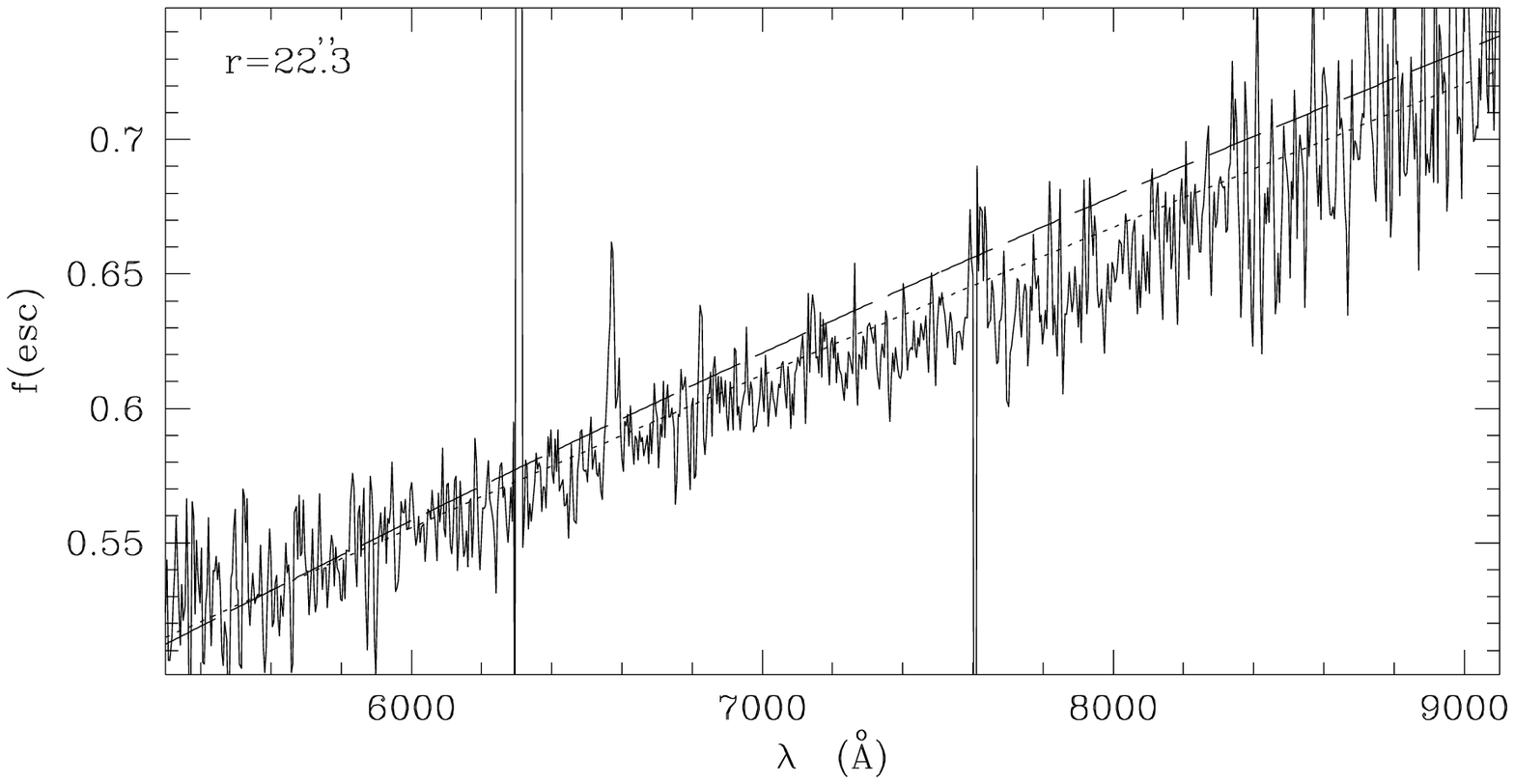}
\caption{Zoom of the ratio of the observed SEDs (solid line) for a pair
of positions symmetrically located at a projected distance $r = 22.3$ arcsec
with respect to the NGC4826 nucleus, on the obscured and unobscured side,
respectively (i.e. the same ratio reproduced in Fig. 2m).
A best-fit WG00 model curve of this ratio, representing $f_{\lambda}(esc)$
for MW-type dust (dotted line) is shown together with a MW extinction curve
(long-dashed line) of the same optical depth (see text).\label{fig3}}
\end{figure}

\clearpage 

\begin{figure}
\plotone{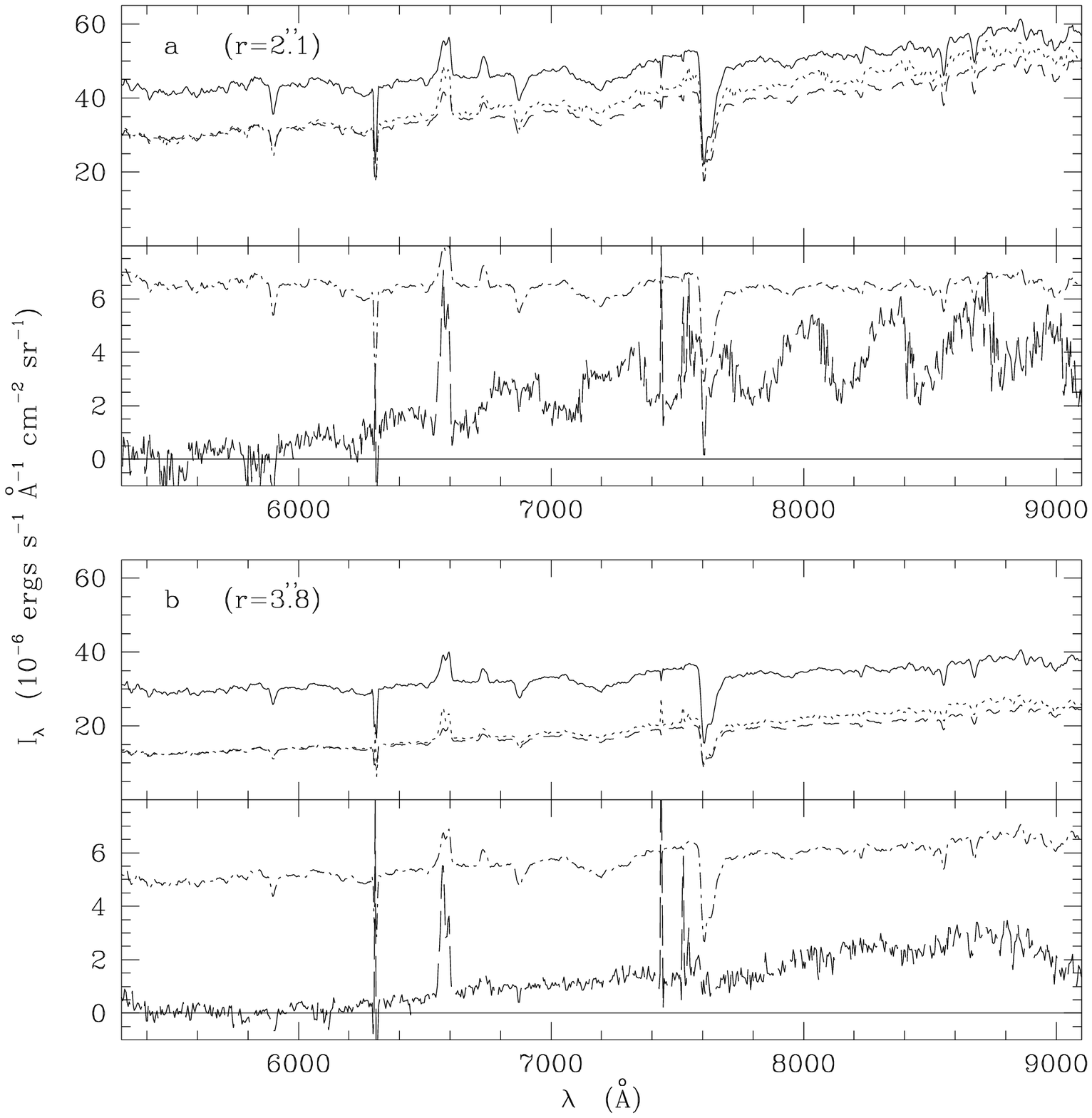}
\caption{The flux density distributions of the intrinsic
(i.e. before dust extinction and scattering) ISRF (solid line),
of the radiation emerging through the dust lane (dotted line)
and of the attenuated ISRF, (short-dashed line), for the same values
of $r$ given in Fig. 2a--n, are shown in the upper panel.
The flux density distributions of the scattered light (dot-short dashed line)
and of the ERE emission (long-dashed line) at $r$ are given
in the lower panel.
For any given value of $r$, the intrinsic ISRF and the radiation emerging
through the dust lane are derived from observation (see text).
The attenuated ISRF and the light scattered by the dust in the Evil Eye
are derived from the intrinsic ISRF of NGC4826 through best-fit modeling
of the ratios reported in Fig. 2a--n (see text).
Finally, the ERE flux density distribution is determined as the residual
of the radiation emerging through the dust lane, after subtraction of
the attenuated ISRF.\label{fig4}}
\end{figure}

\clearpage 

\begin{figure}
\figurenum{4}
\plotone{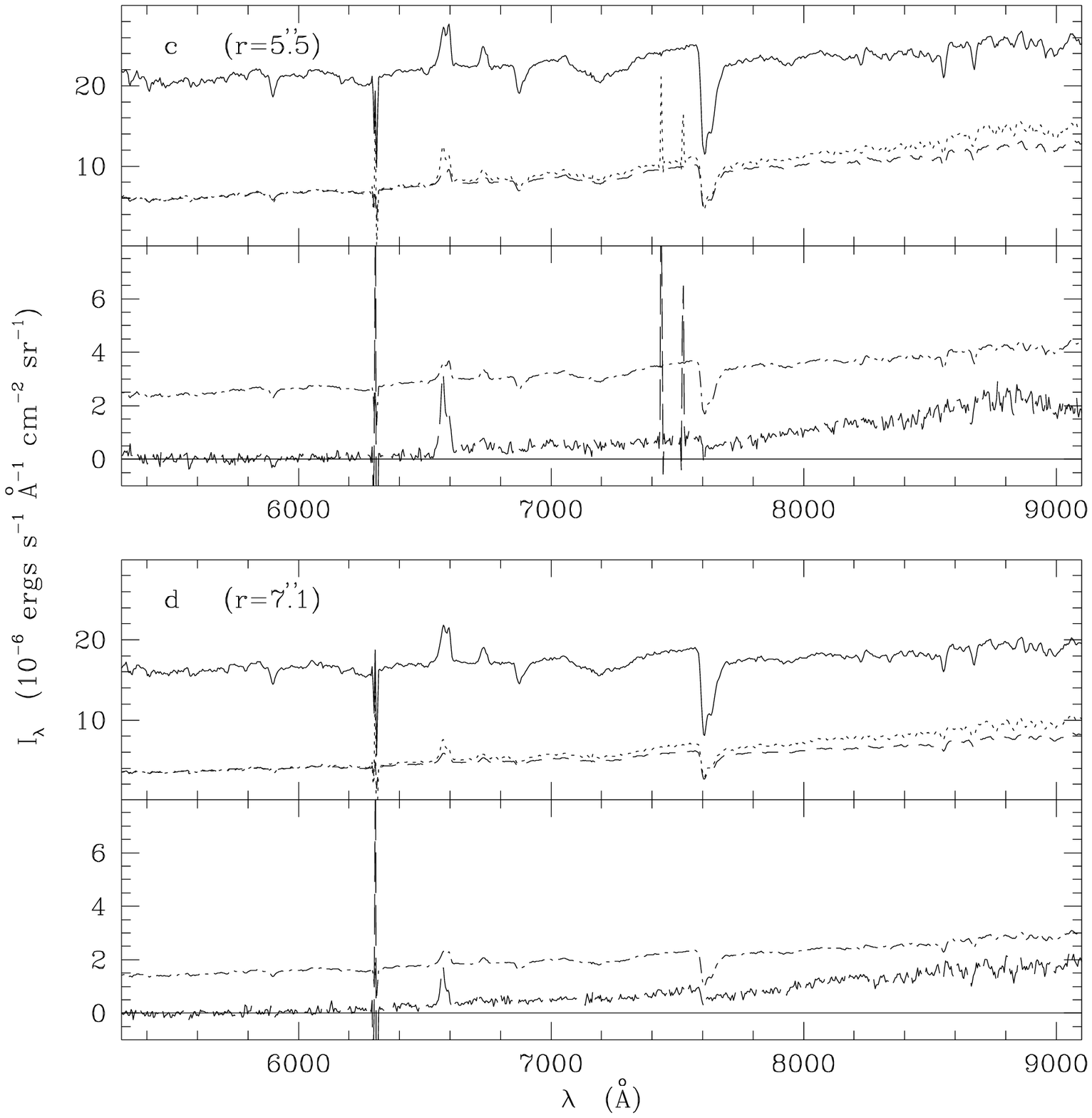}
\caption{continue}
\end{figure}

\clearpage 

\begin{figure}
\figurenum{4}
\plotone{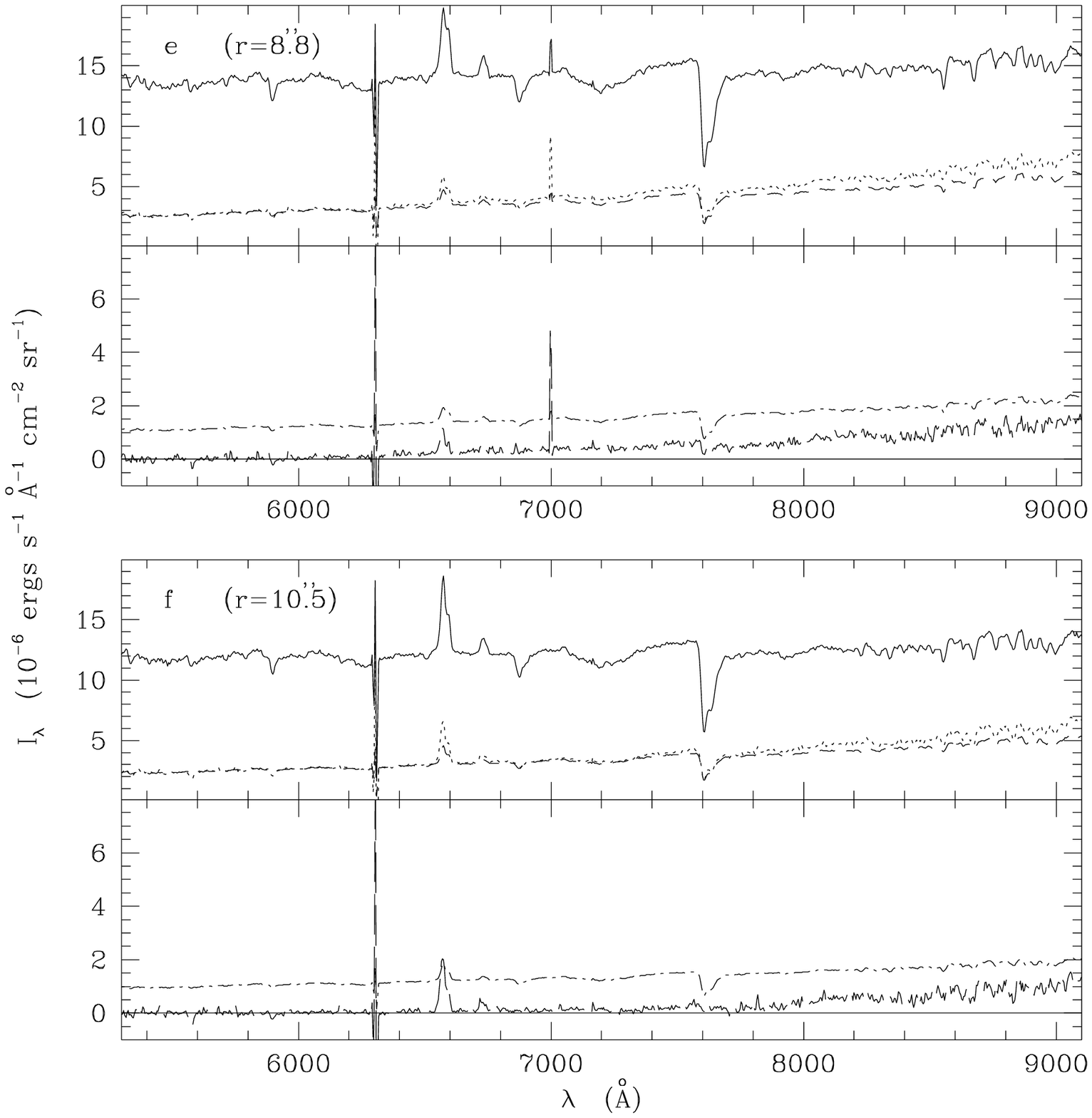}
\caption{continue}
\end{figure}

\clearpage 

\begin{figure}
\figurenum{4}
\plotone{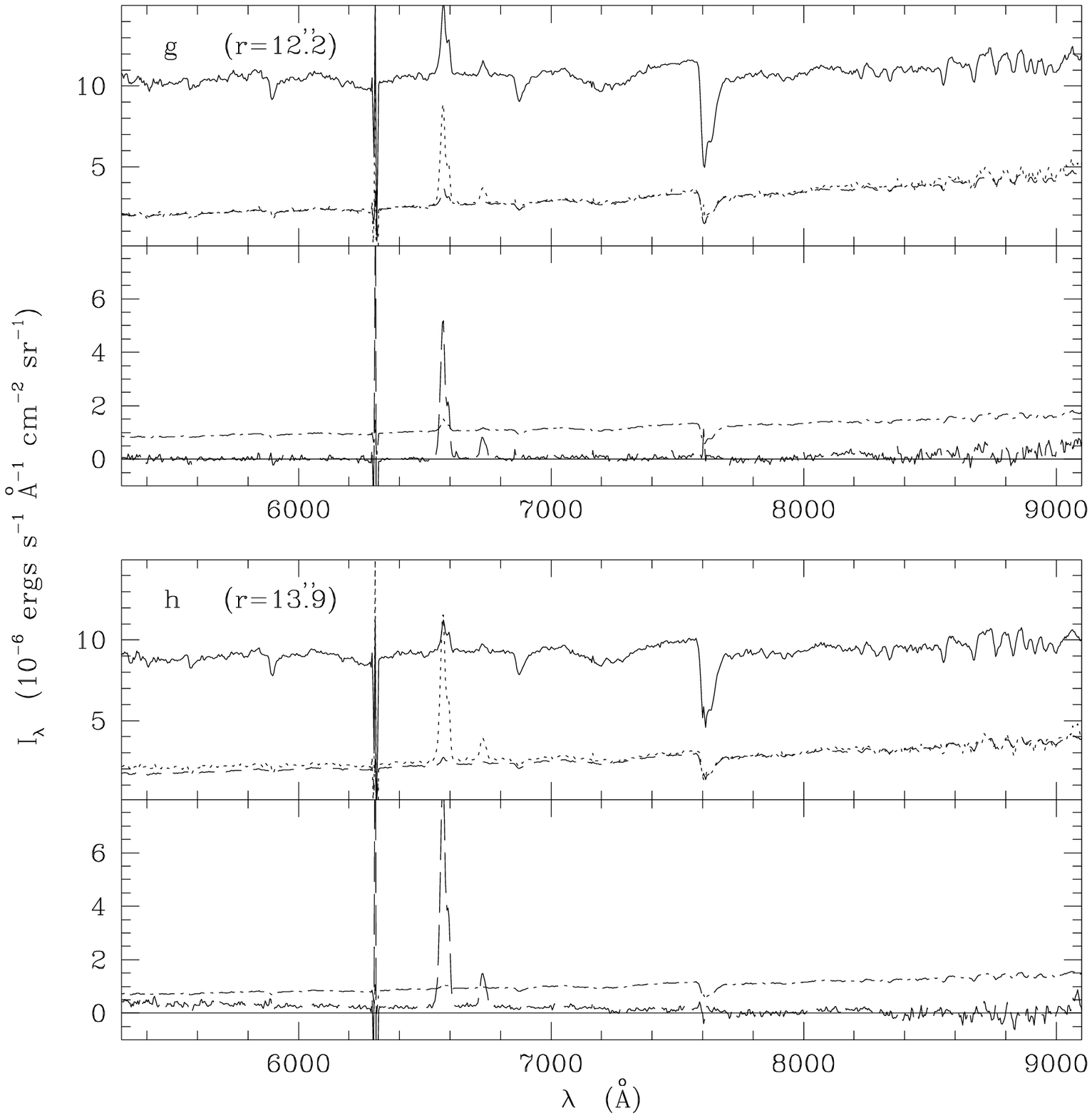}
\caption{continue}
\end{figure}

\clearpage 

\begin{figure}
\figurenum{4}
\plotone{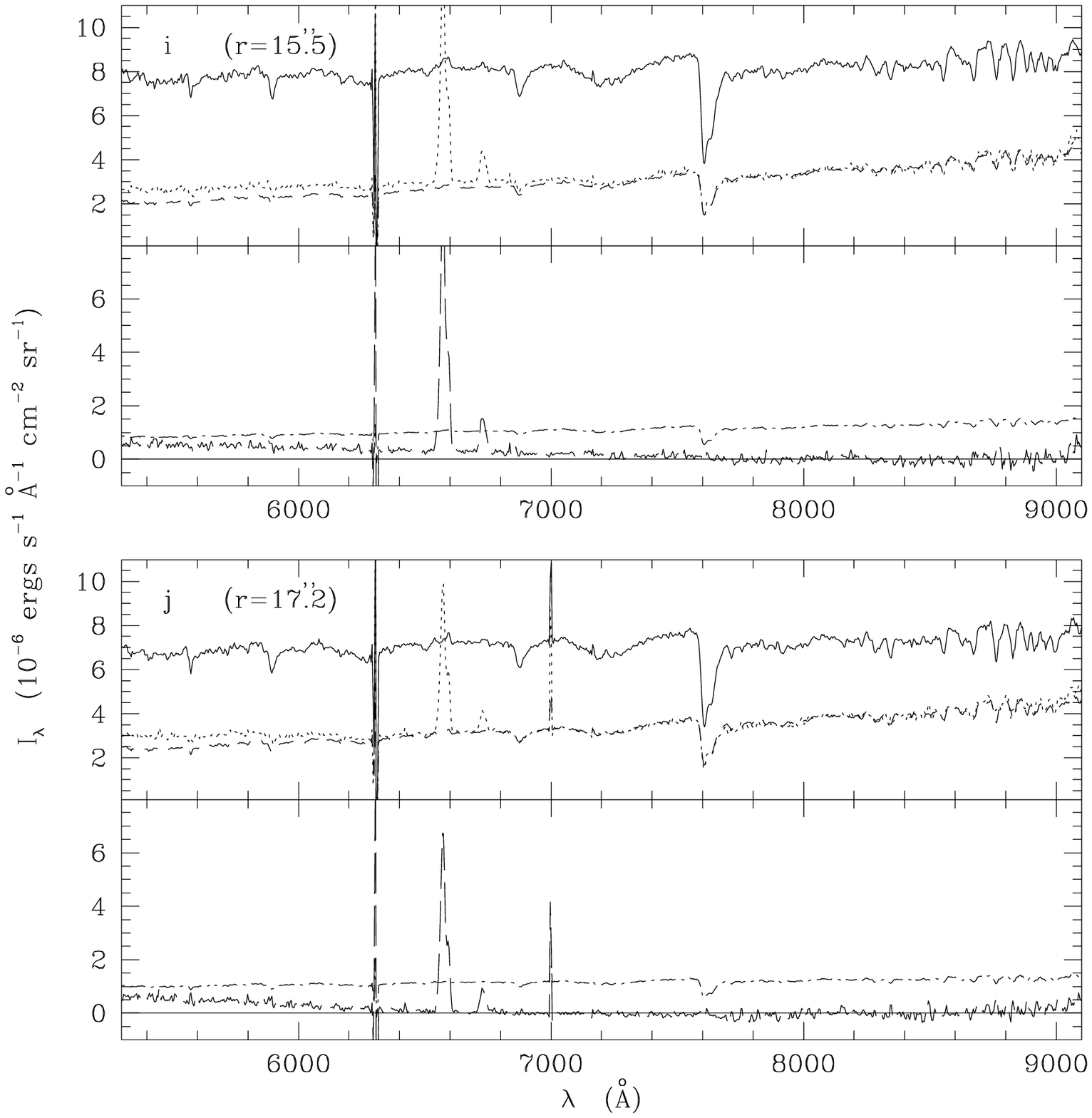}
\caption{continue}
\end{figure}

\clearpage 

\begin{figure}
\figurenum{4}
\plotone{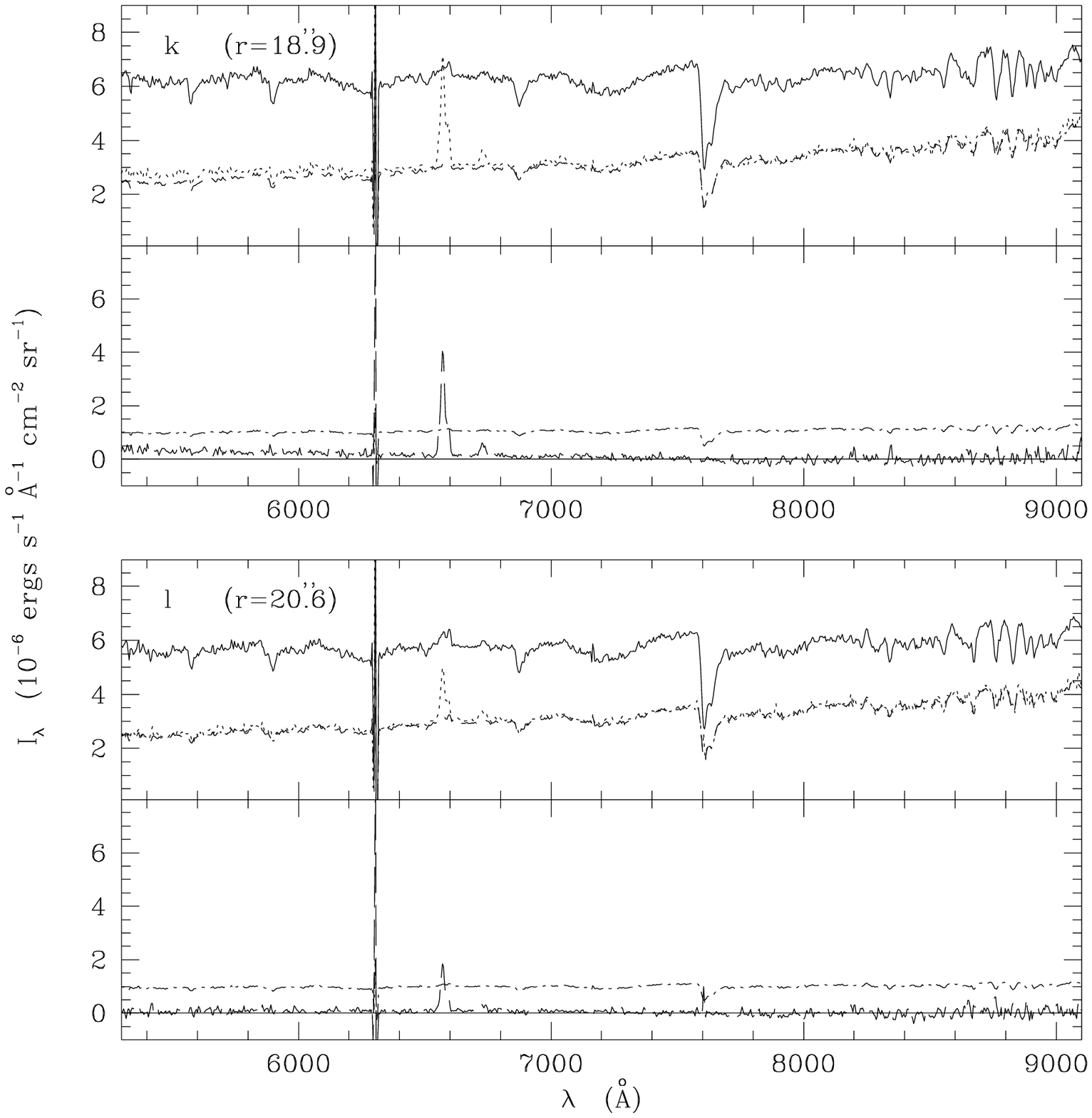}
\caption{continue}
\end{figure}

\clearpage 

\begin{figure}
\figurenum{4}
\plotone{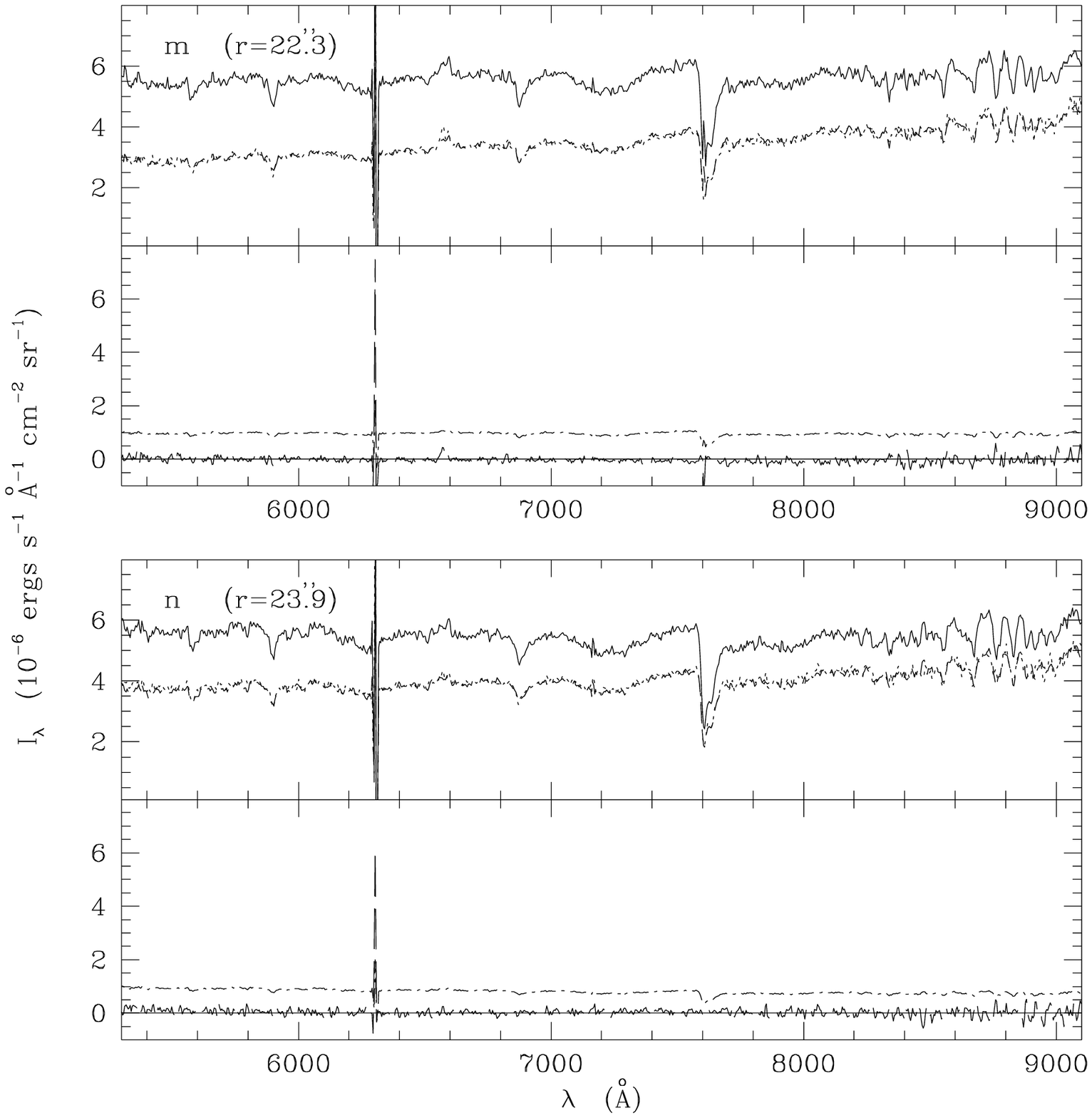}
\caption{continue}
\end{figure}

\clearpage 

\begin{figure}
\plotone{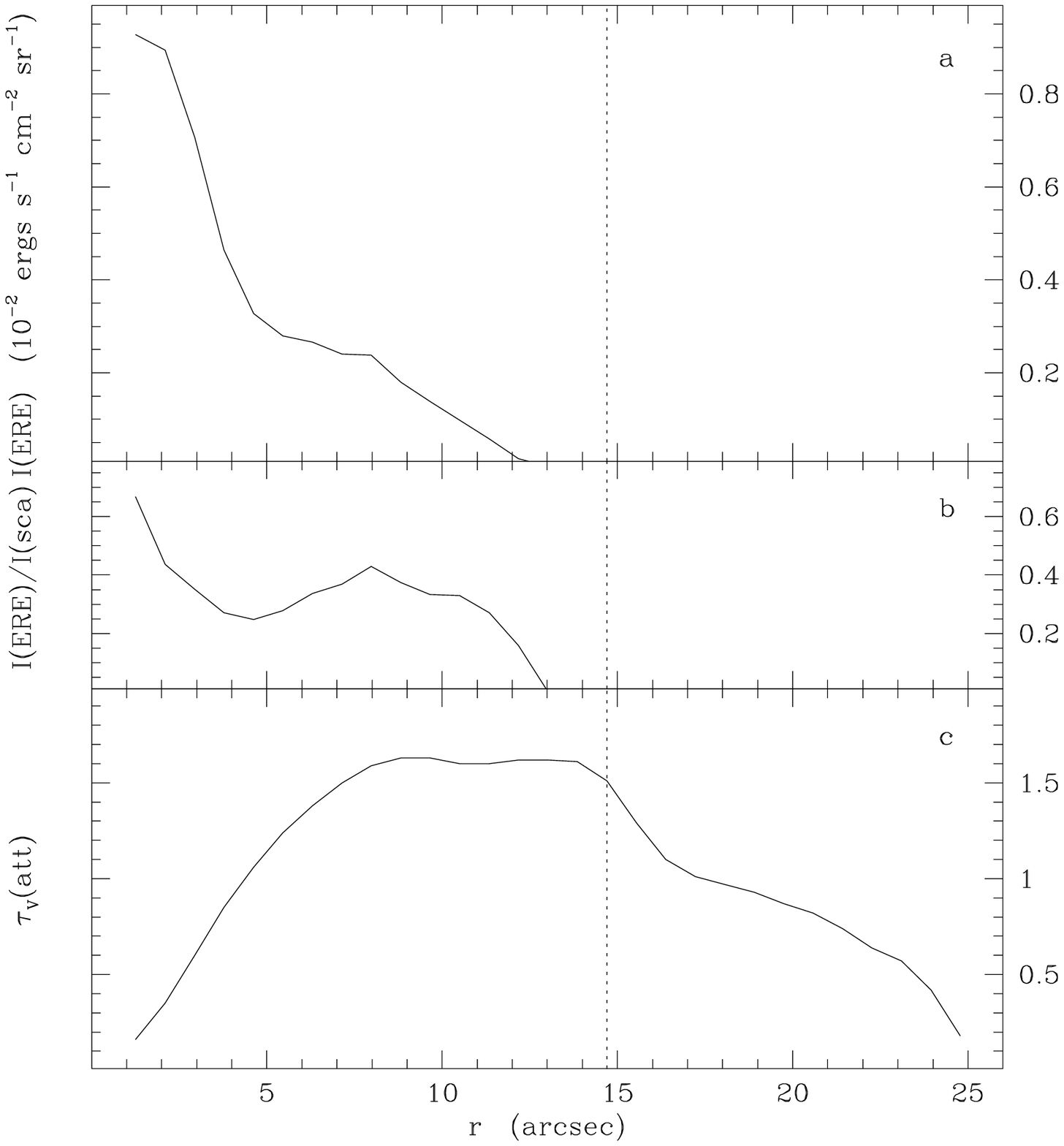}
\caption{The radial distributions of the ERE band-integrated intensity
$I(ERE)$ (a), of the ERE-to-scattered band-integrated intensity ratio
$I(ERE)/I(sca)$ (b), and of the attenuation optical depth in the V-band
$\tau_V(att)$ (c), derived as described in the text, across the dust lane.
A dotted line marks the position of the observed $\rm H \alpha$
($\rm \lambda = 6563~\AA$) line emission peak, corresponding with the broad,
bright HII region intercepted by the spectrograph slit
(cf. Fig. 1 and 7).\label{fig5}}
\end{figure}

\clearpage 

\begin{figure}
\plotone{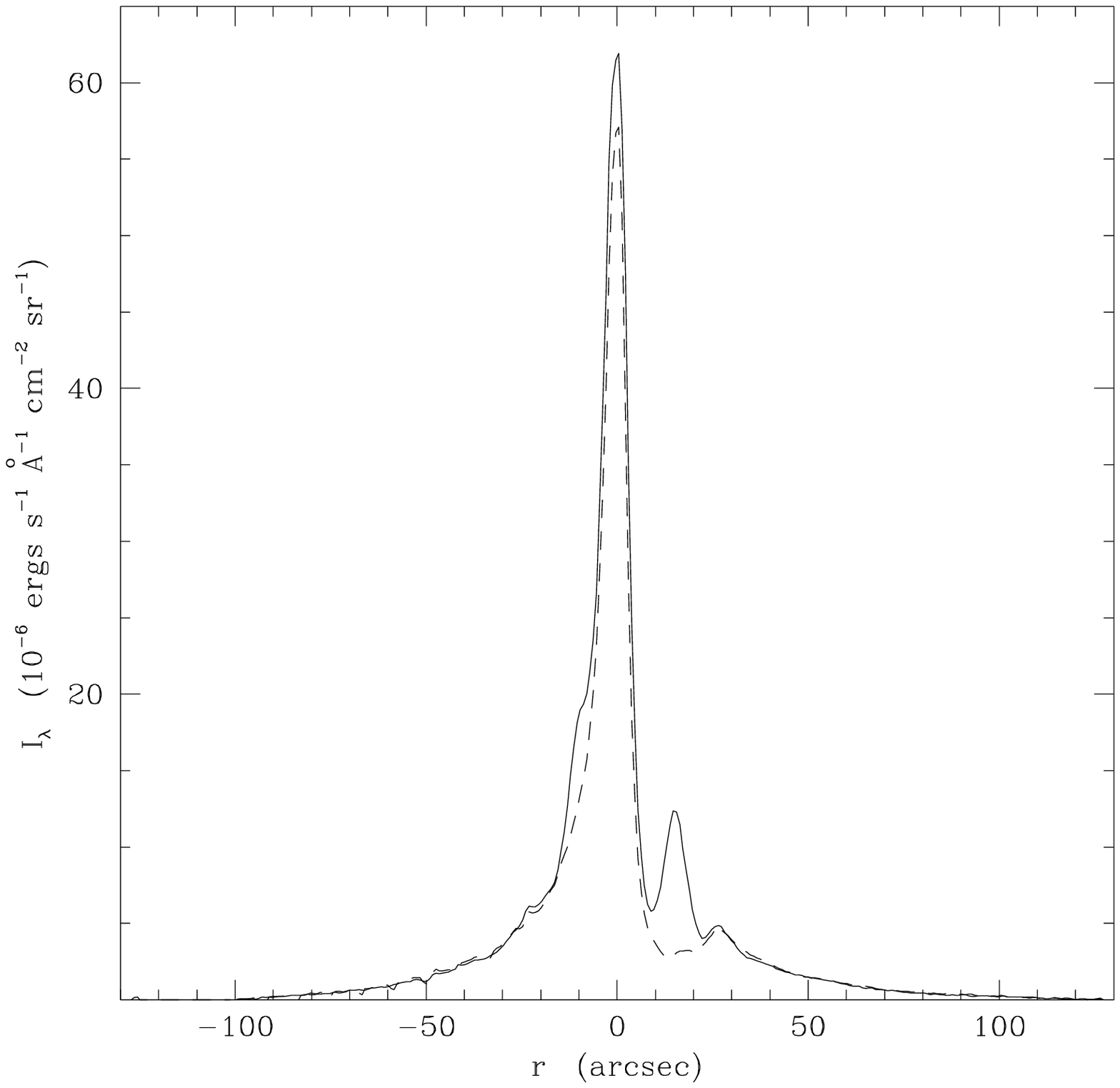}
\caption{Radial profile of the flux density $I_{\lambda}$
at the rest-frame effective wavelength of the R-band
($\rm \lambda = 7000~\AA$ -- short-dashed line)
and at the rest-frame $\rm H \alpha$ line central wavelength
(solid line).\label{fig6}}
\end{figure}

\clearpage 

\begin{figure}
\plotone{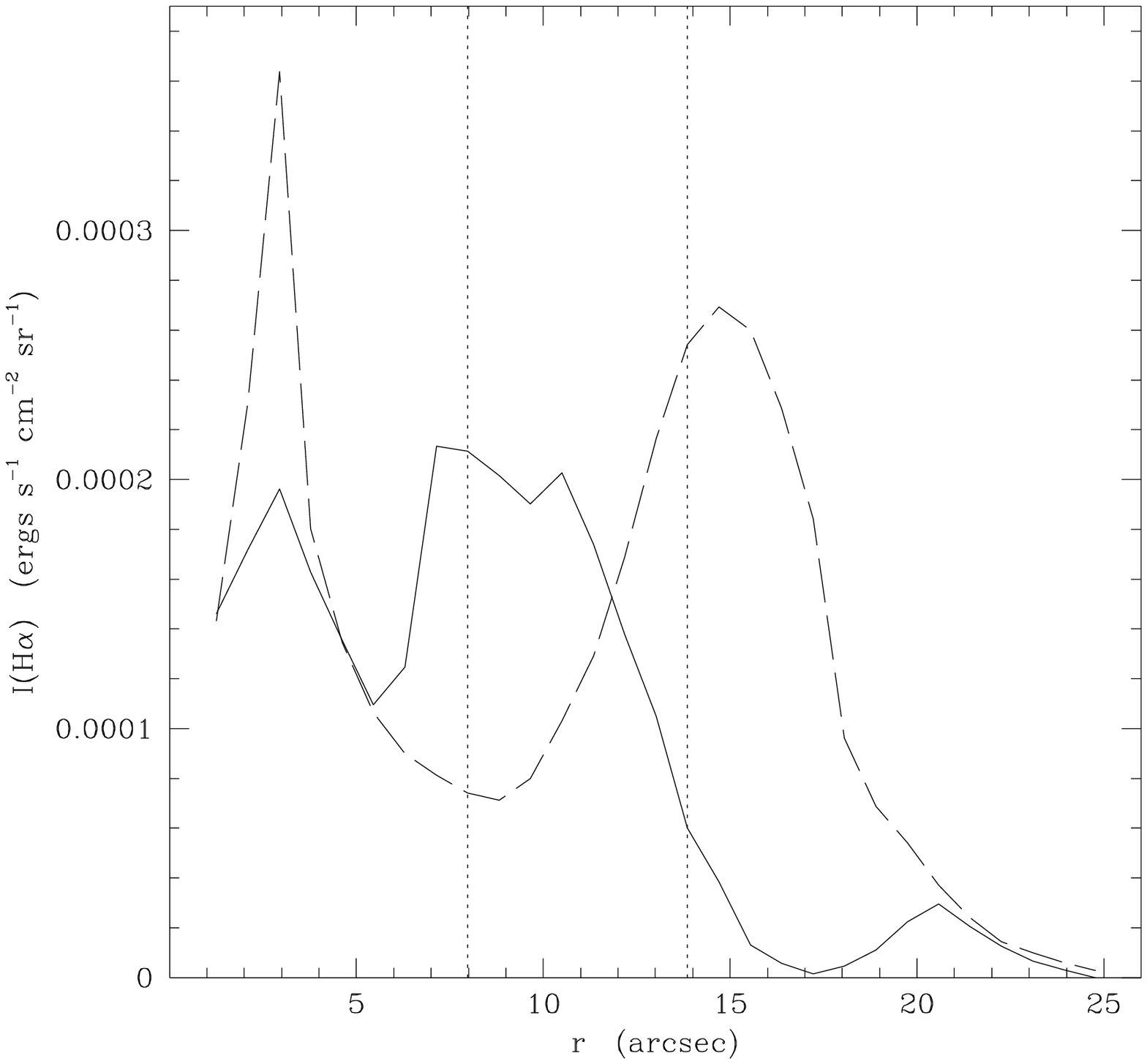}
\caption{Folded radial profile of the $\rm H \alpha$ line intensity
for the dust lane region (long-dashed line) and for an equally extended
region on the opposite side, with respect to the NGC4826 nucleus
(solid line).
The two dotted lines define the region of maximum optical depth across
the dust lane (cf. Fig. 5).\label{fig7}}
\end{figure}

\clearpage 

\begin{figure}
\plotone{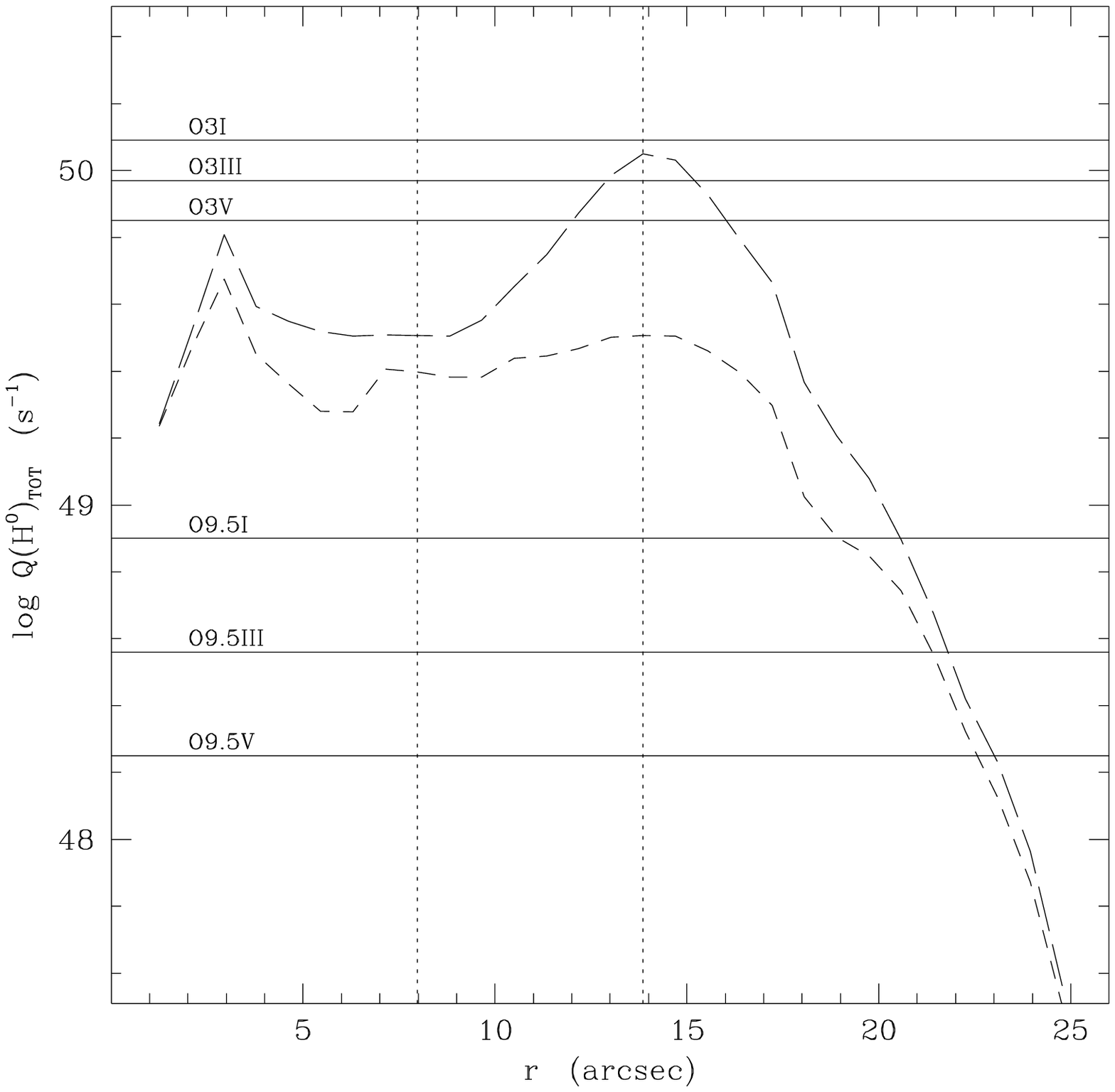}
\caption{Radial variation of the total integrated (shortward of
$\rm 912~\AA$) Lyman continuum (Lyc) photon flux, $Q(H^0)_{TOT}$, corrected
for absorption and scattering, as seen by the observer, across the dust lane.
Two limiting cases for the topology of the sources of
excess (asymmetric) $\rm H \alpha$ line emission, associated with
the Evil Eye, are reproduced, under the assumption that these regions
lie either below the dust lane (case ``a'' -- long-dashed line) or on top
of it (case ``b'' -- short-dashed line).
The integrated Lyc photon flux seen by the Evil Eye is intermediate
with respect to these two estimates (see text).
Values of the integrated Lyc photon flux (at the stellar surface)
for representative individual O-stars of solar metallicity and given
effective temperature $T_{eff}$ and surface gravity $g$ are reproduced
as references (solid lines) from Schaerer \& de Koter (1997).
The two dotted lines define the region of maximum optical depth across
the dust lane (cf. Fig. 5).\label{fig8}}
\end{figure}

\clearpage 

\begin{figure}
\plotone{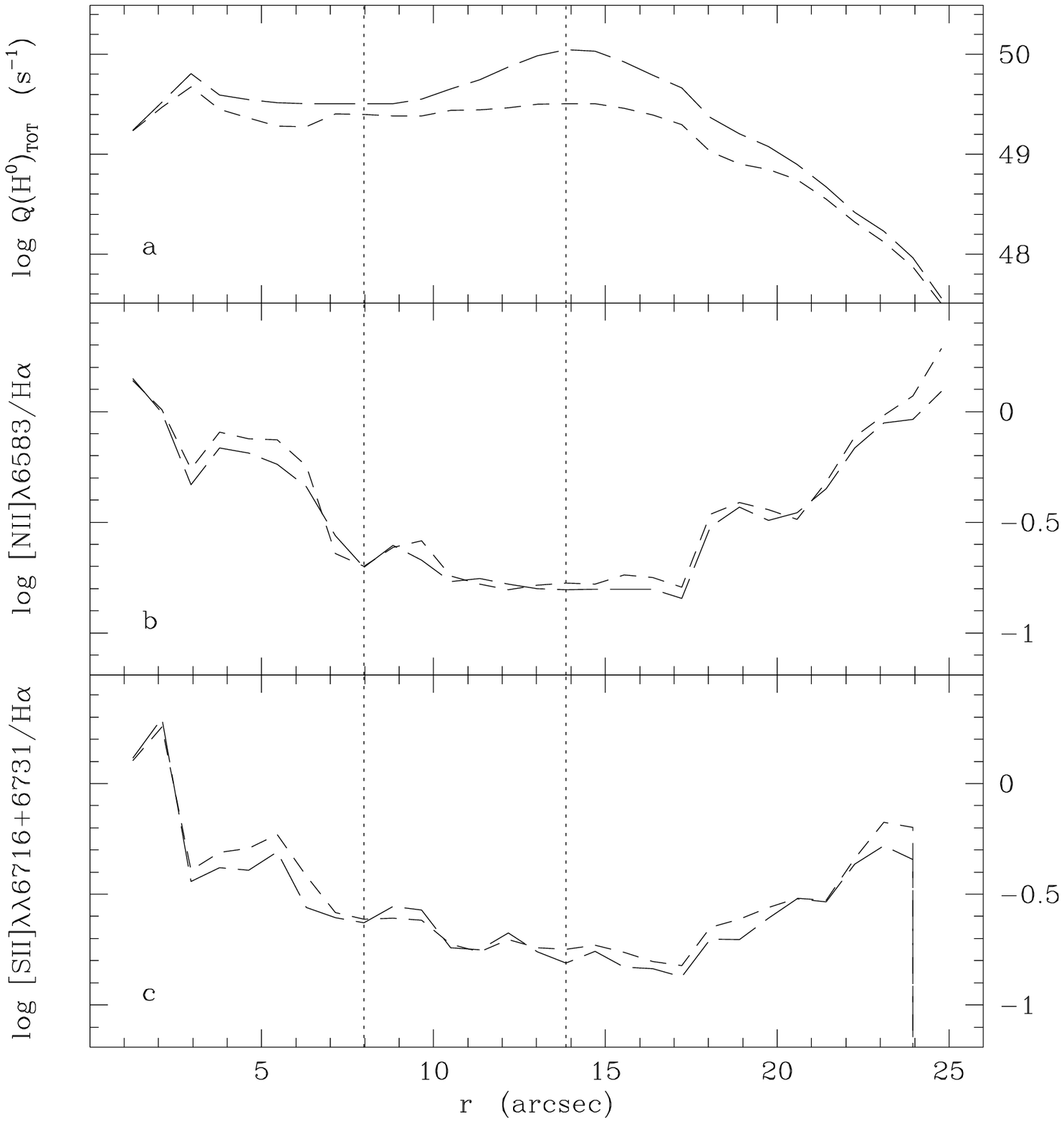}
\caption{Radial variation of $Q(H^0)_{TOT}$ (a), of the ratio
of the [NII] ($\rm \lambda = 6583~\AA$) and the $\rm H \alpha$ line
intensities, $[NII](\lambda 6583)/H \alpha$, (b) and of the intensity
ratio of the [SII] ($\rm \lambda \lambda = 6716, 6731~\AA$) doublet and
the $\rm H \alpha$ line, $[SII](\lambda \lambda 6716+6731)/H \alpha$, (c)
across the dust lane, for the two limiting cases considered in Fig. 8.
The two dotted lines define the region of maximum optical depth across
the dust lane (cf. Fig. 5).\label{fig9}}
\end{figure}

\clearpage 

\begin{figure}
\plotone{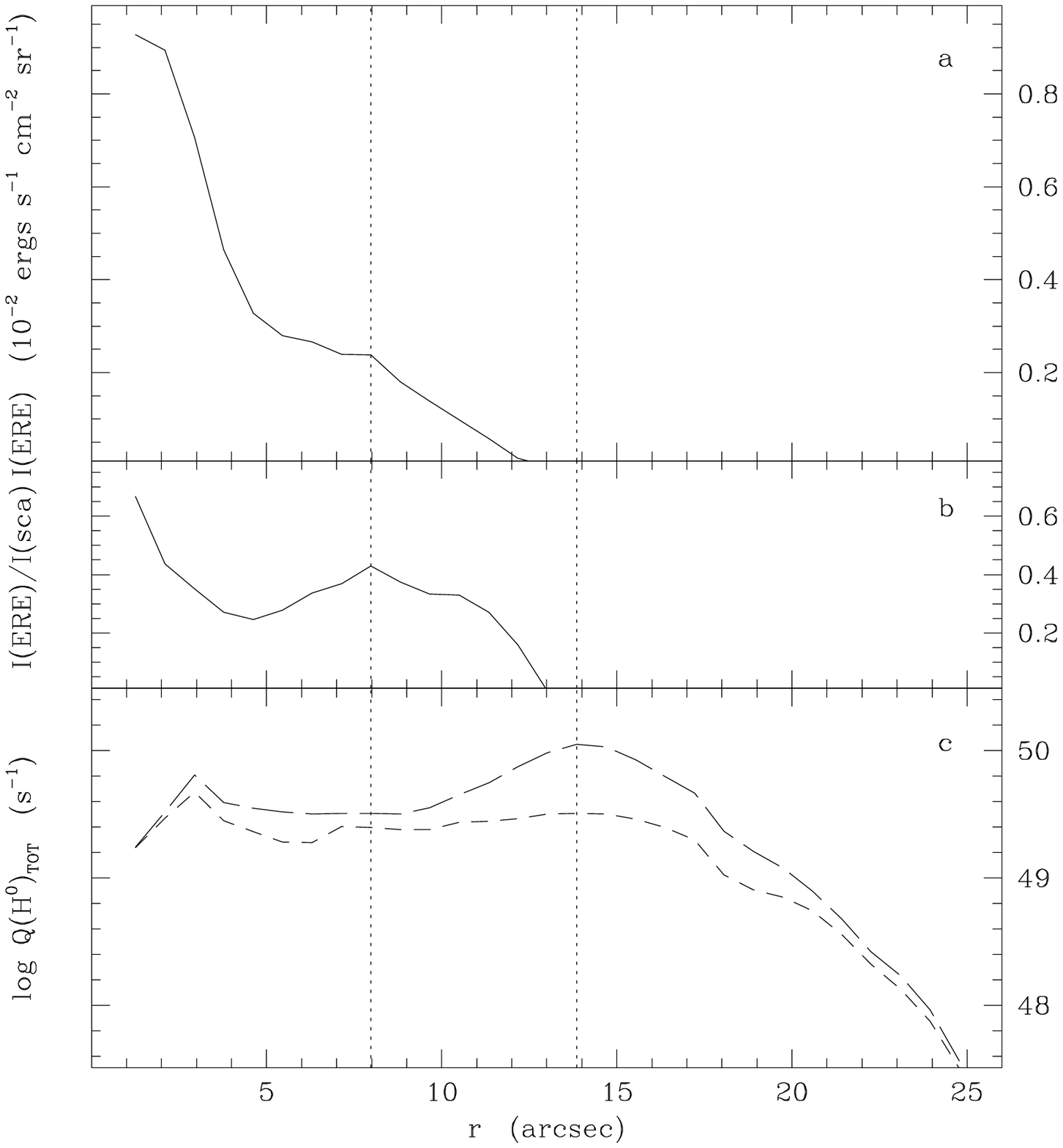}
\caption{The radial distributions of $I(ERE)$ (a), of $I(ERE)/I(sca)$ (b),
and of $Q(H^0)_{TOT}$ (c), for the two limiting cases considered in Fig. 8,
across the dust lane.
The two dotted lines define the region of maximum optical depth across
the dust lane (cf. Fig. 5).\label{fig10}}
\end{figure}

\clearpage 

\begin{figure}
\plotone{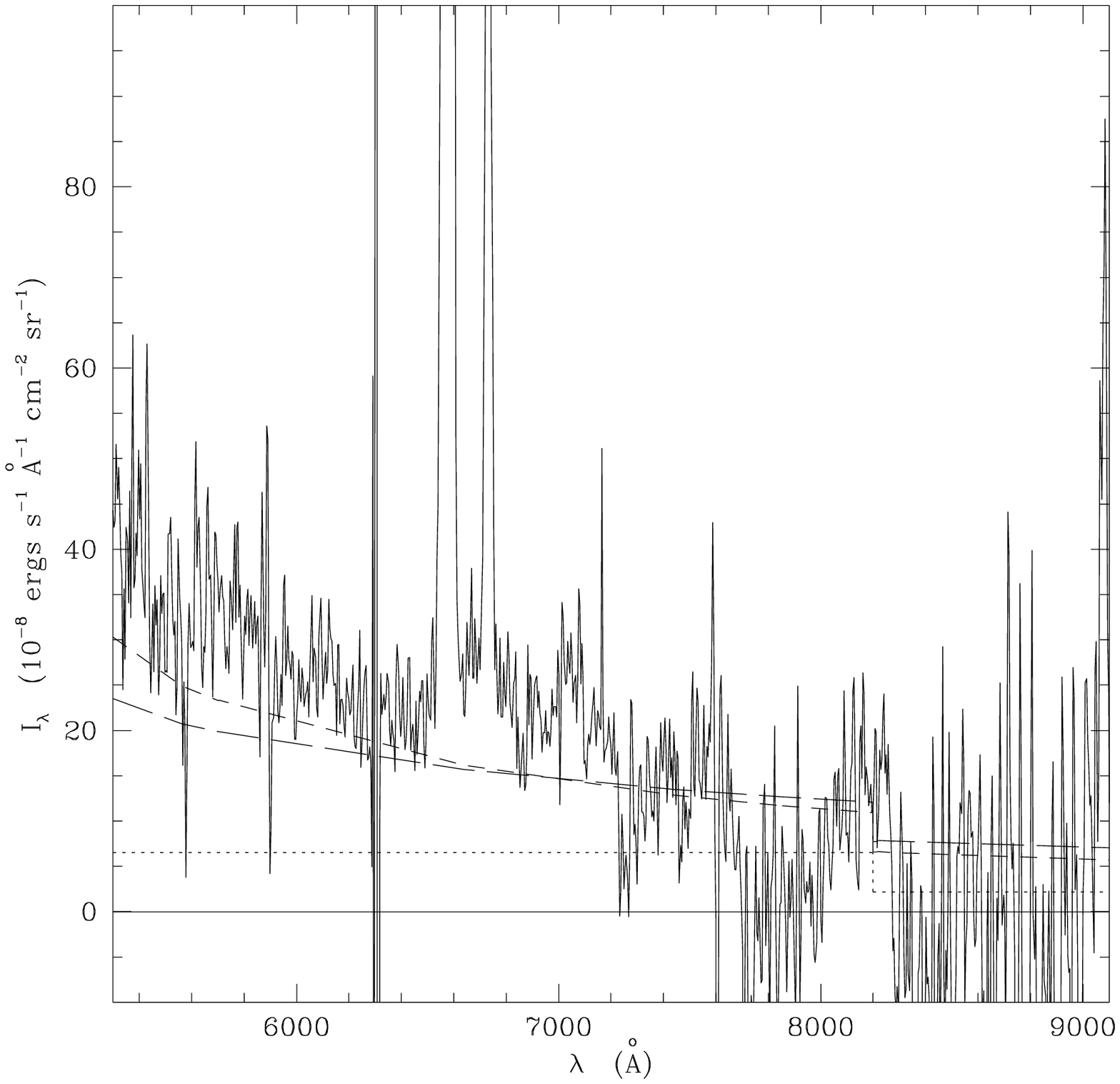}
\caption{Comparison of the predicted profile of the total continuum intensity 
arising from the broad HII region associated with the Evil Eye
and intercepted by our slit with the residual radiation emerging through
the dust lane, after subtraction of the attenuated ISRF, from Fig. 4h.
The predicted total continuum intensity profile is given by
the atomic continuum intensity profile plus the cumulative SED of
the H-ionizing star(s), for the two limiting cases considered in Fig. 8.
The dotted line represents the estimated atomic continuum intensity profile
(see text), for case ``b''.
This represents an upper limit to the observed atomic continuum emission
for case ``a'', so that the predicted total continuum intensity is
an upper limit in this case.\label{fig11}}
\end{figure}

\clearpage

\end{document}